# On the origin and composition of Theia: Constraints from new models of the Giant Impact

M. M. M. Meier*[1,2], A. Reufer[3], R. Wieler[2]

[1]CRPG CNRS Nancy, France
[2]Department of Earth Sciences, ETH Zurich, Switzerland
[3]School of Earth & Space Exploration, Arizona State University, USA

*corresponding author: mmeier@crpg.cnrs-nancy.fr

***Abstract.*** *Knowing the isotopic composition of Theia, the proto-planet which collided with the Earth in the Giant Impact that formed the Moon, could provide interesting insights on the state of homogenization of the inner solar system at the late stages of terrestrial planet formation. We use the known isotopic and modeled chemical compositions of the bulk silicate mantles of Earth and Moon and combine them with different Giant Impact models, to calculate the possible ranges of isotopic composition of Theia in O, Si, Ti, Cr, Zr and W in each model. We compare these ranges to the isotopic composition of carbonaceous chondrites, Mars, and other solar system materials. In the absence of post-impact isotopic re-equilibration, the recently proposed high angular momentum models of the Giant Impact ("impact-fission", Cúk & Stewart, 2012; and "merger", Canup, 2012) allow – by a narrow margin – for a Theia similar to CI-chondrites, and Mars. The "hit-and-run" model (Reufer et al., 2012) allows for a Theia similar to enstatite-chondrites and other Earth-like materials. If the Earth and Moon inherited their different mantle FeO contents from the bulk mantles of the proto-Earth and Theia, the high angular momentum models cannot explain the observed difference. However, both the hit-and-run as well as the classical or "canonical" Giant Impact model naturally explain this difference as the consequence of a simple mixture of two mantles with different FeO. Therefore, the simplest way to reconcile the isotopic similarity, and FeO dissimilarity, of Earth and Moon is a Theia with an Earth-like isotopic composition and a higher (~20%) mantle FeO content.*

# 1. Introduction

*1.1. The Giant Impact and the isotopic conundrum*

The favored model for the formation of the Moon is the "Giant Impact": during the last stage of terrestrial planet formation, planetary embryos of Moon- to Mars-size (approximately 0.01 to 0.1 Earth masses, $M_E$) collide in a sequence of massive mutual collisions – called Giant Impacts – until all or most of them have been accreted into a few, large rocky planets (e.g., Weidenschilling, 2000). In the solar system, two "fully grown" rocky planets – Venus and Earth – formed, while two likely planetary embryos – Mercury and Mars – remained on stranded orbits (e.g., Hansen, 2009). A planetary embryo having between one fourth to two Mars masses (depending on the Giant Impact model; see below), called "Theia" after the mythological mother of the Greek Moon-goddess Selene, then collides with the almost fully formed, differentiated proto-Earth, a few 10 Ma after the formation of the solar system. Such a Giant Impact can explain the observed physical properties of the Earth-Moon system: the high angular momentum, the iron-deficiency of the bulk Moon and the high mass ratio between the Moon and Earth (Hartmann & Davis, 1975; Cameron & Ward, 1976).

In the classical (or "canonical") Giant Impact simulations, most of the mass of the Moon is derived from Theia (e.g., Benz et al., 1986; 1989; Canup & Asphaug, 2001). Therefore, the Moon should chemically and isotopically reflect Theia. Recently, Hosono et al. (2013) and Karato (2014) have pointed out possible problems with the smoothed particle hydrodynamics (SPH) method that is used for most Giant Impact simulations, and have suggested that taking into account these problems might increase the fraction of proto-Earth-derived material in the Moon. In the absence of detailed simulations that quantify the extent of this effect, we will have to work with Giant Impact simulations as published in the recent literature.

While there are some chemical differences between the Earth and the Moon (e.g., the Moon has higher FeO relative to the terrestrial mantle, and is depleted in both volatile and siderophile elements relative to Earth; Jones & Palme, 2000), the Earth and Moon are isotopically identical in all elements which show substantial mass-independent stable isotope variations in meteorites, i.e. O, Ti, Cr and Zr (Wiechert, 2001; Zhang et al., 2012; Qin et al., 2010; Akram, 2013). A small deviation (~12 ppm) in the $\Delta^{17}O$ (see SOM for a definition of that notation) of Earth and Moon has recently been reported, which might either be a signature of Theia, or perhaps of a carbonaceous chondritic late veneer (Herwartz et al., 2014). The Earth and Moon also have an identical Si isotopic composition, which is however (mass-dependently) fractionated towards heavier isotopes, relative to all other solar system materials (e.g., Georg et al., 2007; Fitoussi et al., 2009; Zambardi et al., 2013).

Earth and Moon also have an identical Hf/W ratio, though within relatively large uncertainties (König et al., 2011), and a $\varepsilon^{182}W$ value that is indistinguishable (within uncertainties) from the terrestrial value, at least after correction for the contribution of the late veneer (Touboul et al., 2007; 2009; 2014; Willbold et al., 2011; Kleine et al., 2014). Hence, we are confronted with the "isotopic conundrum" that Theia must have been isotopically more Earth-like than any known (non-lunar) solar system material. Can, or should we expect Theia to be isotopically similar to the Earth? Pahlevan & Stevenson (2007) suggested that the Earth-Moon isotopic similarity is the result of isotopic re-equilibration between the terrestrial magma ocean and the circum-terrestrial disk formed in the Giant Impact, via a common silicate vapor atmosphere. Unfortunately, this elegant solution to the isotopic conundrum seems to be problematic for several reasons (e.g., Melosh, 2010; Salmon & Canup, 2012; Pahlevan & Stevenson, 2012; Nakajima and Stevenson, 2014). We will therefore neglect the effects of such re-equilibration for the purpose of this article.

In the last few years, another possible solution to the isotopic conundrum has been put forward. New variants, or models, of the Giant Impact have been proposed, which predict a significantly lower fraction of Theia-derived material in the Moon than what was thought possible before. Reufer et al. (2012) presented the *hit-and-run model*, where a more massive (0.2 $M_E$) Theia is partially disrupted by the collision, such that about half of its mass escapes to a heliocentric orbit after the Giant Impact. The resulting Moon-forming disk is not as enriched in Theia-derived material as in the canonical model. The angular momentum of the Earth-Moon system in the hit-and-run model was intentionally kept on the order of no more than 10-50% larger than its present value, and is thus relatively low compared to the high-angular momentum Giant Impact models discussed next. Canup (2012) extended the hit-and-run parameter space to somewhat higher masses and the correspondingly higher angular momentum, but much of the parameter space remains unexplored. Reufer et al. (2012) also introduced the parameter $\delta f_T$, to compare the relative abundances of the proto-Earth-derived (silicate) fractions in the Moon and the Earth's mantle. The parameter $f_T$ is the fraction of "target" (proto-Earth) silicate material in the Earth ($f_{TE}$) and the Moon ($f_{TM}$), and $\delta f_T$ = ($f_{TM}$ / $f_{TE}$ − 1) × 100%. Thus, a $\delta f_T$ value of -100% indicates that the Moon is derived exclusively from Theia, while a $\delta f_T$ value of 0% indicates that the Moon has exactly the same proto-Earth-derived (or Theia-derived) fraction as the Earth. We use this parameter here because it allows us to compare different Giant Impact models in a single graph, irrespective of absolute Theia-derived fractions in Earth and Moon. These fractions vary strongly between different Giant Impact models, but for the isotopic composition of Earth and Moon only the *deviation* of the two fractions from one-another is important, and it is this deviation which is expressed by $\delta f_T$. In the hit-and-run model, the typical $\delta f_T$ value is about -35%, a significant improvement over the canonical model ($\delta f_T$ = -70 to -90%). Negative values for $\delta f_T$ dominate for most Giant Impact model runs, although in a few cases positive values have been observed, almost exclusively within the model from Canup (2012; see below). For simplicity, and as in Canup (2012), we will always discuss the absolute $\delta f_T$ values (= $|\delta f_T|$), so that *low* $|\delta f_T|$ values indicate very *similar* Theia-derived fractions of Earth and Moon, while *high* $|\delta f_T|$ values correspond to *dissimilar* fractions of Theia-derived material in Earth and Moon.

A second new Giant Impact model, which we will call the *impact-fission model*, was presented by Cúk & Stewart (2012). It is the first dynamical implementation of the earlier idea of impact fission (e.g., Wänke & Dreibus, 1986). Here the proto-Earth is spinning rapidly before the Giant Impact. Theia is smaller than in the canonical model, between 2.6 and 10% of the Earth's mass. Its impact leads to a massive ejection of (partially vaporized) Earth mantle material into a circum-planetary disk. As we will show below, the typical outcome of that model is a $\delta f_T$ value of about -8% (Figure 1). The angular momentum of the Earth-Moon system produced in this kind of Giant Impact is considerably higher than its present-day value (up to a factor of 2.5), and excess angular momentum needs to be lost via the "evection resonance" between the Earth, the Moon and the Sun (Cúk & Stewart, 2012; Touma & Wisdom, 1998). A third new Giant Impact model has been proposed by Canup (2012). In this model, which we will call the *merger model*, two large planetary embryos with a mass ratio of 0.6:0.4 or 0.55:0.45 merge to form the Earth and its Moon. Since the two planetary embryo masses are almost equal, their relative contributions to the Moon-forming disk are also nearly equal, leading occasionally to $\delta f_T$ values near 0% (Figure 1). Again, the resulting angular momentum of the Earth-Moon system is much larger than observed today, and the merger model therefore also requires the evection resonance mechanism to be compatible with the present-day Earth-Moon system. For this reason, we will sometimes refer to both the impact-fission and merger models collectively as the *high angular momentum models*.

The new Giant Impact models potentially solve the isotopic conundrum. However, the authors of the new models have usually not discussed in detail the possible range of isotopic compositions that are now possible, or "allowed", for Theia. Which solar system materials are now possible precursors for Theia? What can we learn about Theia's possible origins, or the isotopic heterogeneity between different planetary embryos in the closing phase of terrestrial planet formation? It is the main purpose of this work to study the possibilities opened by the new Giant Impact models. For reasons that will become apparent below, we will focus our analysis on three types of solar system material as a possible proxy for the isotopic composition of Theia: carbonaceous chondrites, Mars, and materials that we will call "Earth-like". The latter are (groups of) meteorites that are either very similar, or even indistinguishable from the Earth in many (or all) isotopic systems, including, e.g., the enstatite chondrites, the enstatite achondrites (aubrites) and some ungrouped achondrites introduced below.

**2. Methods**

*2.1. Data sources for chemical and isotopic compositions*

In all Giant Impact simulations, the cores of the proto-Earth and Theia merge quickly (within hours) and are thus not expected to equilibrate with the silicates (e.g., Canup & Asphaug, 2001). Therefore, we consider only the silicate part of Earth, Moon and Theia when constructing the mixing equations. This is important for Cr and W where a relatively high fraction of the bulk abundance is in the core, and for Si, where incorporation into the core results in mass-dependent isotopic fractionation. For our mass-balance calculations, we use the values of McDonough & Sun (1995) for the composition of the bulk silicate Earth. The abundance of the bulk silicate Moon in O, Si, Ti, Cr, Zr and W was assumed to be identical to the corresponding values of the bulk silicate Earth. For both Earth and Moon, relative uncertainties in chemical composition of 5% for O, 10% for Si, Ti and Zr, 15% for Cr and 20% for W were adopted, as suggested by McDonough & Sun (1995). The resulting range of possible lunar abundances of these elements encompasses e.g. the compositional model of the Moon by O'Neill (1991) and the W concentration of the Moon as calculated by Touboul et al. (2007).

For the isotopic composition of different solar system materials (Earth, Moon, and meteorites), we used the values given by Herwartz et al. (2014), Wiechert et al. (2004), Franchi et al. (1999), Clayton & Mayeda (1996; 1999), Clayton et al. (1991), Shukolyukov et al. (2010) for $\Delta^{17}$O; Zambardi et al. (2013), Armytage et al. (2011; 2012) for $\delta^{30}$Si; Zhang et al. (2012), Trinquier et al. (2009) and Leya et al. (2008) for $\varepsilon^{50}$Ti; Qin et al. (2010a; 2010b), Trinquier et al. (2007), Shukolyukov & Lugmair (2006), Yamakawa et al. (2010), Shukolyukov et al. (2010) for $^{54}$Cr; Akram (2013), Akram & Schönbächler (2014) for $^{96}$Zr, and Touboul et al. (2007; 2009; 2014), Kleine et al. (2009; 2014) and Willbold et al. (2011) for $^{182}$W. All values used are given in supplementary table S1. The mass balance equations we used to construct Figures 3-7 are also derived and explained in the supplementary materials. For our calculation of the possible ranges of isotopic composition for Theia, we al-ways assume that no post-impact isotopic re-equilibration between the Earth's magma ocean and the Moon-forming disk, as proposed by Pahlevan & Stevenson (2007), takes place.

*2.2. Selection of "successful" Giant Impact simulations*

Only a subset of the simulation runs of the Giant Impact (in all four models) yields an Earth-Moon system that matches the observed one in general physical properties. We define a subset of Giant Impact simulation runs from the literature (Reufer et al. 2012, Cúk & Stewart, 2012 and Canup, 2012) that we call "successful". In successful simulation runs, the resulting satellite has a total complement of silicates of 67-150% of a lunar mass after accretion (Kokubo, 2000), and a metallic iron/silicate ratio of less than 5% (given by the inferred radius of the lunar core of <350 km; Weber et al., 2011). Table S2 in the supplementary materials contains the parameters of the outcomes of all successful simulation runs of the three new models considered here, as well as the averages for each of the three new models.

**3. Results & Discussion**

*3.1. General results*

Figure 1 shows the "successful" (solid symbols) and "unsuccessful" (open symbols) Giant Impact simulation runs from the recent literature, allowing us to identify typical outcomes of the different models. Figure 2 explains the motivation for testing a carbonaceous chondritic origin for Theia (as discussed below). Figures 3 – 7 show the results of our mass-balance calculations for the range of possible isotopic compositions of Theia. They clearly illustrate the "isotopic conundrum" of the formation of the Moon: it is very difficult to find a solar system proxy material that could serve as an analogue for Theia, at least for the very high values of $|\delta f_T|$ observed in the canonical Giant Impact. In particular, it is very difficult to form Theia from carbonaceous chondrites, which show very large deviations from the isotopic composition of the Earth. Except for Earth-like materials (including enstatite meteorites), very low values of $|\delta f_T|$, about 10% and below, are required for possible matches between Theia and known solar system materials. While the hit-and-run model by Reufer et al. (2012) provides a gradual improvement over the canonical model (e.g., Canup & Asphaug, 2001), the high angular momentum models (impact-fission and merger) with their even lower $|\delta f_T|$ values extending below 10% are much more promising in resolving the isotopic conundrum by means of a lower $|\delta f_T|$ value. However, as we will show below, $|\delta f_T|$ is not the only relevant parameter, and other aspects of the models (e.g., the mass of Theia, or its FeO content) might play an important role. We

will now first look at the general outcomes of the new Giant Impact models. Then, we will discuss three possible proxy materials for Theia: carbonaceous chondrites, Mars, and Earth-like materials. We will discuss the fractionation of the Si isotopes of Earth and Moon relative to the other solar system materials, the interesting problem of the different FeO contents of the Earth and Moon and the implications this might have for Theia and the Giant Impact. We summarize our results in table 1, which is intended to serve as reference for future work in this field.

*3.2. Typical outcomes of the four Giant Impact models*

Before we can look into the possible range of isotopic compositions of Theia, we have to determine the typical outcome of each of the three new Giant Impact models. For the canonical Giant Impact model, no $|\delta f_T|$ values were given in the literature at the time, but they can be reconstructed from the fraction of Theia-derived material in the Moon, the mass of Theia, and the pre- and post-Giant impact masses of the Earth, which are usually given (e.g., from Canup & Asphaug, 2001). A Theia-derived fraction of 80% in the Moon, and mass of Theia of about 0.1 $M_E$ translates into a $|\delta f_T|$ value of 77% (see SOM), and a plausible range of $|\delta f_T|$ for the canonical model is about 80±8%, based on literature values (Canup & Asphaug, 2001; Canup, 2004; Canup, 2008). As shown in Figure 1, the various simulation runs of the three new Moon-forming Giant Impact models show a large scatter in both the resulting mass of the Moon (ordinate) and the value of $|\delta f_T|$ (abscissa). There is no overall trend, although at least within the hit-and-run model (Reufer et al., 2012; hit-and-run simulation runs from Canup, 2012 are also included), there is a weak correlation between more massive satellites and lower $|\delta f_T|$ values. In all models, only a subset of the simulation runs result in a satellite with a mass (67-150%) and iron/silicate ratio (<0.05; the remainder being too massive, not massive enough, or too rich in iron) corresponding to the actual Moon. For the hit-and-run model, 9 out of 62 runs reported by Reufer et al. (2012), and 5 out of 65 runs reported by Canup (2012) are "successful" as defined above, or ~11% overall. For the impact-fission model, 19 out of 85 (22%) reported runs are successful (Cúk & Stewart, 2012), the same fraction as the 11 out of 49 (22%) successful runs of the merger model (Canup, 2012). The successful Giant Impact simulations from all three new models are represented by solid symbols in Figure 1, and their parameters are summarized in supplementary table S2. For the hit-and-run model, the successful simulation runs cluster roughly at $|\delta f_T| = 30 - 40\%$, right about where the weak correlation for that model reaches ~1 Moon-mass. The successful hit-and-run Giant Impact simulations have an average $|\delta f_T|$ of 39%, a standard deviation of 11%, and a full range (lowest to highest reported) from $|\delta f_T| = 23\%$ to 67% (this range has been used in Figures 3 – 7).

Values of $|\delta f_T|$ smaller than ~20% are restricted to the high-angular momentum models. The successful runs of the impact-fission model (Cúk & Stewart, 2012) cluster at about $|\delta f_T|$ = 5 – 10%. The average $|\delta f_T|$ value for this model is 8% with a standard deviation of 5%, and a range spanning from $|\delta f_T|$ = 4 to 20%, as shown in Figures 3 – 7. Both the hit-and-run and the impact-fission model are quite predictive about the distribution of Theia- vs. proto-Earth derived material in the Earth and Moon, as their $|\delta f_T|$ values cluster relatively tightly in Figure 1. On the other hand, the successful simulation runs from the merger model (Canup, 2012) show a wide scatter from $\delta f_T$ = -35% to +37% (positive $\delta f_T$ values indicating rare cases of a Moon dominated by proto-Earth material), with no clusters. This large scatter results in an average $|\delta f_T|$ value of 19% with a standard deviation of 13% for the merger model. The full range of $|\delta f_T|$ values for that model extends from 0.3% to 37%, and is shown in Figures 3 – 7. While $\delta f_T$ values around 0% are certainly possible in the merger model (two out of the twelve successful merger runs yield $\delta f_T$ = -1% and -0.3%, respectively), they can certainly not be considered typical outcomes.

### 3.3. A Theia similar to carbonaceous chondrites?

Carbonaceous and non-carbonaceous materials[*] follow clearly separated trends in their mass-independent anomalies in $\Delta^{17}O$, $\varepsilon^{50}Ti$ and $\varepsilon^{54}Cr$ (Figure 2; Warren, 2011; see SOM for a definition of the $\Delta$ and $\varepsilon$ notations). This suggests an origin in separate reservoirs, perhaps outer and inner solar system (Warren, 2011; Rubin & Wasson, 1995). While the Earth belongs to the "non-carbonaceous" field in Figure 2, its data point in this figure plots at the carbonaceous-facing end of the field. Therefore, it could be said that the Earth is more enriched in carbonaceous material than any other material falling into the non-carbonaceous field. When mixing LL chondrites with an average carbonaceous chondrite (close to CM or CR), the Earth would contain about 30-40% carbonaceous material, 10-20% more than Mars (Warren, 2011; Fitoussi & Bourdon, 2012). Could this "excess" have been added by a 0.1-0.2 $M_E$ Theia of carbonaceous origin? The "carbonaceous" Eagle Station pallasites and the "carbonaceous" basaltic achondrite NWA 011 (Bogdanovski & Lugmair, 2004; Trinquier et al., 2009) show that at least some differentiated objects of carbonaceous origin must have existed in the early solar system. The Moon's chemical composition, including its excess in FeO relative to the terrestrial mantle, is reasonably well explained by a carbonaceous Theia contributing ~20% to the mass of the Moon (O'Neill, 1991). The Earth's pattern of volatile element abundance is most similar to the one of carbonaceous chondrites (Allègre et al., 2001; Palme & O'Neill, 2003). About 13% of the Earth's mass had to be added at a late stage by volatile-rich material, to satisfy constraints from short-lived radionuclides (Schönbächler et al., 2010), a similar fraction to the one added by Theia in the Giant Impact. Using Figures 3 – 7, we therefore first discuss whether Theia might have been of carbonaceous origin, i.e., whether it could have belonged to the "carbonaceous field" shown in Figure 2. In Figure 3, we show the possible range of isotopic composition of Theia in $\Delta^{17}O$ (between the two black solid lines), as a function of the $|\delta f_T|$ value of the Giant Impact that formed the Moon. The lower $|\delta f_T|$, the larger this range becomes, accommodating more and more known solar system materials as it decreases towards zero. The ranges of $|\delta f_T|$ for the three new Giant Impact models as discussed above are shown in shaded and hatched areas (see caption of Figure 3 for more details). Figure 3 shows that a CI-chondritic composition in $\Delta^{17}O$ (=0.38±0.09; Clayton & Mayeda, 1999; Table S1) for Theia is possible only if $|\delta f_T|<6\%$, and thus restricted to some simulation runs from the two high angular momentum models. All other types of carbonaceous chondrites are excluded for $|\delta f_T| > 1\%$. Figures 4 – 6 have the same basic design as Figure 3, but consider the isotope systems of $\varepsilon^{50}Ti$, $\varepsilon^{54}Cr$ and $\varepsilon^{96}Zr$ instead. In Figures 4 and 5, it is obvious that a CI-chondritic composition for Theia is again possible only under the high-angular momentum models, with all other carbonaceous chondrites possible only for very low values of $|\delta f_T|$. This would suggest that if Theia was of carbonaceous origin, it must most likely have been related to (or accreted from) CI chondrites. Interestingly, O'Neill (1991) proposes a chemical model of the Moon, where a CI-chondritic Theia contributes about 20% of the mass of the Moon. However, the corresponding $|\delta f_T|$ of ~18% is incompatible with observational constraints (Figures 3 – 5), except for $\varepsilon^{96}Zr$ (Figure 6). A CI-chondritic Theia under a high angular momentum scenario with a $|\delta f_T| = 4\%$ (at the lower end of the impact fission range) is possible by a narrow margin, and would suggest that the Moon should likely have nominally slightly higher $\Delta^{17}O$, $\varepsilon^{50}Ti$ and $\varepsilon^{54}Cr$ values than the Earth. However, at such low $|\delta f_T|$, the resulting isotopic composition of the Moon is also very sensitive to the assumed bulk silicate O, Ti and Cr content of the Earth and Moon, which is not known precisely enough to make an accurate prediction for the isotopic composition of the Moon in that case.

---

[*]*Here we use the term "carbonaceous" as it has been used historically, i.e., referring to a certain class of solar system materials rather than to the carbon content of the material.*

We conclude this section with a note on the "icy Theia" model that Reufer et al. (2012) presented. This variant of the hit-and-run model involves an ice-rich Theia with a differentiated mantle consisting of 50% water ice, similar to objects in the outer solar system, e.g., the largest jovian satellite, Ganymede, or the dwarf planet Ceres (Thomas et al., 2005; Küppers et al., 2014). Such objects might have been introduced into the inner solar system during Giant planet migrations (Walsh et al., 2011; O'Brien et al., 2014), and it has even been proposed that Ceres is such a "displaced" object (e.g., McKinnon, 2008; 2012). The original motivation for the "icy Theia" model was the hope that a large fraction of the mantle water would, although at first still gravitationally bound in the circum-terrestrial disk, escape the system after the Giant Impact (e.g., through hydrodynamic escape), leaving behind a disk strongly enriched in debris derived from the proto-Earth, thereby explaining the isotopic conundrum. But while some of the "icy" simulation runs reached $\delta f_T$ close to 0% in the hit-and-run model, the resulting mass of the silicate disk for these runs was far too low to yield the Moon (Figure 1). Nevertheless, the coverage of parameter space for this model is far from complete. With the remote possibility of a CI chondritic Theia, we suggest that the "icy Theia" model should be revisited, e.g., in the context of the high angular momentum models, in particular the impact fission model by Cúk & Stewart (2012).

In summary of this section, a CI-chondritic Theia is the only possible proxy material from the "carbonaceous field" of solar system materials, possible by a narrow margin under a few exceptional runs of the high angular momentum models. However, this CI-chondritic Theia is not compatible with the one proposed by the O'Neill (1991) model, as it contributes less material to the bulk Moon than required by that model. The higher fraction of carbonaceous material in the Earth (e.g., relative to Mars; Warren, 2011), and the 13% of volatile-rich matter added to the Earth in late accretion (Schönbächler et al., 2010) can thus not have been added in the Moon-forming impact, but must have an earlier origin.

*3.4. A Mars-like Theia?*

Mars is the only other planet besides the Earth from which we have samples (in the form of meteorites). It has about the mass of a planetary embryo, and has thus often been suggested as an isotopic proxy for Theia (e.g., Pahlevan & Stevenson, 2007; Zhang et al., 2012). Dauphas & Pourmand (2011) show that Mars must have accreted within 2 Ma after the formation of the solar system, and suggest that Mars might be a surviving planetary embryo. This view is supported by n-body simulations of planetary accretion, which have difficulties explaining the low mass of Mars unless it is a planetary embryo stranded in an otherwise depleted region of the planet forming disk (e.g., Hansen, 2009). In this view, Mars did not form in-*situ*, but was dynamically ejected from the vicinity of the Earth at an early stage. An early isolated Mars might not be strictly representative of the planetary embryos which accreted onto Earth in the following ~100 Ma. Nevertheless, given it is the only other planet sampled, and perhaps even a planetary embryo, it is worthwhile to determine if a Theia might have been Mars-like. Schönbächler et al. (2010) also suggested a Theia with a Mars-like complement of volatiles could have delivered the Earth's volatiles.

A Mars-like Theia does not necessarily have to have exactly the same isotopic composition as Mars. Different planetary embryos might have had different isotopic compositions, depending on where exactly they formed, and what types of material they incorporated during accretion. It has repeatedly been suggested that there were isotopic trends present in the planet forming disk, e.g. for O isotopes (Pahlevan & Stevenson, 2007; Belbruno & Gott; 2005) or for Cr isotopes (Shukolyukov & Lugmair, 2000). Such trends might have led to a variation of isotopic composition between different planetary embryos. Mars and Theia could thus be seen as two possible examples of the typical range of variations that one might expect to be present in the disk. A "Mars-like" Theia thus indicates an object that has an isotopic composition that is similarly different from the Earth as Mars. From Figures 3 and 4, it is obvious that the neither the hit-and-run nor the canonical model can accommodate a Mars-like Theia. Even a Theia with half the deviation in $\Delta^{17}O$ and $\varepsilon^{50}Ti$ of Mars is only possible under the high angular momentum models.

Tungsten is another element that can potentially provide insights into the origin and composition of Theia. The strongly lithophile element Hf has a radioactive isotope, $^{182}Hf$, which decays with a half-life of 8.9 Ma into $^{182}W$. Tungsten (W) is moderately siderophile. The radiogenic excess of $^{182}W$ in the core and mantle of a differentiated planetary body will thus depend on the time of core formation and the Hf/W of the planetary mantle (see Kleine et al., 2009, for a thorough introduction into the subject). In contrast to early reports (see Jones & Palme, 2000, and references therein), the bulk silicate mantles of the Earth and the Moon seem to have, within uncertainties, identical Hf/W ratios (König et al., 2011; Münker, 2010) and $\varepsilon^{182}W$ values after taking into account the likely addition of late veneer material (Willbold et al., 2011; Touboul et al., 2007; 2009; 2014; Kleine et al., 2014; Kleine et al., in preparation). This requires Theia to have an Earth-like Hf/W ratio only if the Theia-derived fractions of the Earth and Moon are very different, i.e., as in the canonical Giant Impact model. If these contributions are similar, as in the more recent Giant Impact models, even large initial differences in Hf/W (as they are expected among planetary embryos, see, e.g., Nimmo et al., 2010) would disappear as Theia and proto-Earth material is mixed inside both Moon and Earth. The same is true for $\varepsilon^{182}W$. In Figure 8, we show the "allowed" pairs of Hf/W and $\varepsilon^{182}W$. Theia could have had a Hf/W and a $\varepsilon^{182}W$ value similar to Mars, under the hit-and-run model. Even higher deviations are possible under the high angular momentum models. Also shown in Figure 8 are the

Meier et al., 2014 – Origin and composition of Theia

Hf/W and $\epsilon^{182}$W values of various planetary embryos from the accretion model by Nimmo et al. (2010). Only a single embryo plots within the bounds of the canonical model, a total of nine plot within the bounds of the hit-and-run model, and the remaining (total 24) all plot within the bounds of the high angular momentum models. Since a very early-formed planetary embryo like Mars (Dauphas & Pourmand, 2011) can be accommodated just as well as a late-formed, Earth-like, planetary embryo, by both the hit-and-run and high angular momentum models (although not the canonical model), the Hf/W and $\epsilon^{182}$W cannot provide strong constraints on the origin, formation and differentiation history of Theia (see also Burkhardt & Dauphas, 2014).

In summary of this section, while the high angular momentum models of the Giant Impact can allow for a Theia of Mars-like isotopic composition in O, Ti, Cr, Zr, Si, the canonical and hit-and-run models cannot. The Hf-W system cannot provide very tight constraints on the origin of Theia, and already the possible range of Hf/W ratios and $\epsilon^{182}$W values resulting from the hit-and-run model is compatible with most of the predicted range of modeled planetary embryos.

*3.5. A Theia from enstatite meteorites or other Earth-like materials?*

Enstatite chondrites and achondrites (aubrites) have an isotopic composition highly similar to the one of the Earth (e.g., $\epsilon^{48}$Ca: Dauphas et al., 2014; $\epsilon^{54}$Cr: Shukolyukov & Lugmair, 2000; $\epsilon^{64}$Ni: Regelous et al., 2008; $\epsilon^{92}$Mo: Burkhardt et al., 2011; Figure 2). This has led to the proposal of enstatite chondrite models of the Earth (e.g., Javoy, 1995; Javoy et al, 2010), which have been challenged by authors referring to the different Mg/Si ratios of the Earth's mantle and enstatite chondrites (Jagoutz et al., 1979; Palme & O'Neill, 2003), and the large mass-dependent fractionation of Si isotopes between the Earth-Moon system and enstatite chondrites (Fitoussi & Bourdon, 2012). The enstatite meteorites are also not a perfect match for the Earth in some isotope systems, e.g., they have slightly higher $\Delta^{17}$O (Herwartz et al., 2014) and $\epsilon^{53}$Cr (Shukolyukov & Lugmair, 2000) as well as slightly lower $\epsilon^{50}$Ti (Zhang et al., 2012) values, and also show a higher deficit in s-process-produced Mo isotopes (see Burkhardt et al., 2011, for details) than the Earth. Even if the Earth itself might not have accreted from enstatite chondrites, we can test whether Theia might have. Fitoussi & Bourdon (2012) give an upper limit of ~15% enstatite chondrite material in the Earth, again a number very similar to the mass fraction added to Earth in the Giant Impact. Also, Herwartz et al. (2014) show that an EH-like Theia would be a good fit to the observed difference in $\Delta^{17}$O between the Earth and the Moon, if the $|\delta f_T|$ was on the order of ~30-40%, i.e., similar to the typical value of the hit-and-run model. The arguably most important difference between Earth and the enstatite chondrites is in the mass-dependent fractionation of Si isotopes (Georg et al., 2007; Fitoussi et al., 2009; Armytage et al., 2011; Zambardi et al., 2013). The Earth and Moon are equal, but distinctively heavier in their Si isotopic composition than all other solar system materials, including enstatite chondrites. This is thought to be a consequence of isotopic fractionation during the incorporation of metallic Si into the Earth's core (e.g., Georg et al., 2007; Fitoussi et al., 2009; Armytage et al., 2011;2012; Zambardi et al., 2013). As Mars has a chondritic Si isotopic composition, the required planetary mass for Si incorporation must be higher than the mass of Mars, i.e., >0.11 $M_E$. Heterogeneous accretion models of the Earth suggest that the incorporation of Si into the core starts soon after the mass of the planet has exceeded 0.1 $M_E$, and reaches a maximum at about 0.2 to 0.3 Earth masses, although the exact path depends on the accreted material and the evolution of the oxidation state of the mantle (Wade & Wood, 2005; Wood et al., 2008). The very light Si isotopic composition of enstatite chondrites (even lighter than other chondrite groups), and the observational upper limit on the Si fraction of the core, excludes them as significant building blocks of the Earth (Fitoussi & Bourdon, 2012). On the other hand, additional fractionation processes for Si are conceivable, and whether or not they played an important role during the formation of the Earth is unclear (Zambardi et al., 2013). The similarity of the Si-isotopic composition of Earth and Moon also seems to constrain the possibility of an enstatite chondritic Theia: Figures 3-6 show that Theia could have had an isotopic composition in $\Delta^{17}$O, $\varepsilon^{50}$Ti, $\varepsilon^{54}$Cr and $\varepsilon^{96}$Zr similar to enstatite meteorites, under each of the Giant Impact models. The crucial problem is shown in Figure 7: No Giant Impact model (except a few merger model runs with $|\delta f_T| < 4\%$) can accommodate the very light Si isotopic composition of an enstatite chondritic Theia.

However, this does not take into account the possibility that Si isotopes might have been fractionated within Theia as well. In both the merger and the hit-and-run models, Theia is more massive than Mars (at 0.45 and 0.20 $M_E$, respectively), i.e., massive enough to start incorporating Si into its own core (Wade & Wood, 2005) and thus fractionate the Si isotopic composition of its mantle. The core of Theia merges immediately with the proto-Earth core in Giant Impact simulations, without significant re-equilibration with the mantle. Therefore, the mantle of a Theia accreted from enstatite chondrites must have been isotopically heavier in $\delta^{30}$Si than bulk enstatite chondrites, for both the merger and hit-and-run model. If the resulting $\delta^{30}$Si value of the theian mantle was roughly similar to the one of the Earth today (or the proto-Earth), this would naturally explain the similarity of Earth and Moon in $\delta^{30}$Si. The fractionation of Si isotopes in the theian mantle is a function of Si incorporation into the theian core. However, if the amount of Si incorporated into the core of Theia is very high, it will dominate the total amount of Si in the Earth's core after the Giant Impact, limiting the amount of Si the proto-Earth can have incorporated, and thus limiting the fractionation of Si isotopes in the proto-Earth mantle. Therefore, the observation of an identical Si isotopic composition of the Earth and Moon also constrains the possible range of Si content of Theia's core. This is shown in Figure 9, where we plot $\delta^{30}Si$ (for proto-Earth, Theia, Earth and the Moon) against the fraction of Si in Theia's core (the model parameters are given in supplementary table S3, and the model is explained in the SOM). For the hit-and-run model (Figure 9a), there is a window of about ~6-20% Si in Theia's core for which the Earth and Moon end up with the same $\delta^{30}Si$ (within uncertainty). Also, the absolute value of $\delta^{30}Si$ within this window is compatible, within uncertainty, with the one measured in terrestrial and lunar rocks. Therefore, the hit-and-run model can accommodate a enstatite chondritic Theia as long as its core Si content was between 6 and 20%. If Theia's core had more than 10% Si, the Moon becomes nominally heavier in $\delta^{30}Si$ than the Earth. Georg et al. (2007) reported lunar samples having a slightly heavier Si isotopic composition than the Earth, but this was not confirmed by any other authors working on Si isotopes (e.g., Fitoussi et al., 2009; Armytage et al., 2011; Zambardi et al., 2013). For the merger model simulation run with the lowest $|\delta f_T| = 0.3\%$ (Figure 9b), the isotopic compositions of the Earth and Moon are indistinguishable irrespective of the theian core Si fraction, and an enstatite chondritic Theia is possible as long as Theia had less than ~18% Si in its core.

Another type of extraterrestrial, isotopically Earth-like material (for most isotopes) is represented by the ungrouped primitive achondrites NWA 5400 (and paired stones; Greenwood et al., 2012; Day et al., 2012; Shukolyukov et al., 2010) and NWA 4741 (Garvie et al., 2012). These meteorites are indistinguishable from the Earth in their $\Delta^{17}O$. For NWA 4741, no isotopic data beyond O isotopes have been reported. NWA 5400 also has the same $\varepsilon^{54}Cr$ value as the Earth, albeit also a somewhat higher $\varepsilon^{53}Cr$ (+0.5$\varepsilon$). While its $^{53}Mn$-$^{53}Cr$-age of <4541 Ma (Shukloyukov et al., 2010) suggests a late formation, the difference in (initial) $\varepsilon^{53}Cr$ also proves that it is not a piece of the Earth (i.e., a "terrene meteorite"). This is also supported by the high Pb-Pb age (Amelin & Irving, 2011). However, given the high concentration of highly siderophile elements, NWA 5400 is certainly not a piece of Theia or another large, differentiated body, either (Shukolyukov et al., 2010). Nevertheless, NWA 5400 and other primitive achondrites with $\Delta^{17}O$ values indistinguishable from the one of the Earth (e.g., NWA 4741; Garvie et al., 2012), as well as enstatite chondrites and achondrites, might represent samples from an inner disk uniform reservoir (IDUR; Dauphas et al., 2014) that extended out to about 1.5 AU (but did not include Mars). Most of the material from IDUR would have been accreted onto Earth and Venus, and would thus not be common in the asteroid belt, i.e., among the parent bodies of meteorites falling to Earth today. Nevertheless, some of that material might have become implanted in the asteroid belt and might still occasionally be sampled by processes delivering meteorites to Earth. Note that even within the most Earth-like materials, there is some residual heterogeneity in isotopic composition (e.g., $\varepsilon^{53}$Cr in NWA 5400). The high $\varepsilon^{53}$Cr of NWA 5400 precludes a Theia made of this material under the canonical model, but all other Giant Impact models are compatible. It would be of particular interest to study the Si isotopic composition of NWA 5400, to see if all Earth-like materials are related to enstatite chondrites and their very light $\delta^{30}$Si, or if some Earth-like materials with chondritic $\delta^{30}$Si (~-0.4 to -0.5) exist.

In summary of this section, both the hit-and-run and merger models can accommodate a Theia accreted from enstatite chondrites, by fractionating some Si (6-20%, and <18%, respectively) into the theian core, but the impact fission and canonical models cannot. The hit-and-run model predicts that there should be a difference in the Si isotopic composition of Earth and Moon if the theian core mass fraction of Si deviates strongly from ~10%. Earth-like materials are compatible with all three new Giant Impact models – although not with the canonical model due to $\varepsilon^{53}$Cr – and are thus interesting candidate materials for Theia. However, they need further analysis, and in particular their Si isotopic composition should be very revealing.

*3.6. The difference in the FeO content of Earth and Moon*

Contrasting with the isotopic similarity of the Earth and Moon are some important differences in chemical composition. This includes a depletion, relative to the Earth, in volatile and siderophile elements in lunar rocks (e.g., Jones & Palme, 2000). The origin of these depletions is currently unclear. Incorporation of siderophile elements during core formation on the newly formed Moon has been suggested (Jones & Palme, 2000; O'Neill, 1991). The loss of lithophile volatiles (e.g. alkali elements) during the Giant Impact seems unlikely given the fact that e.g. K isotopes are not fractionated accordingly (Humayun & Clayton, 1995). If the low volatile element concentrations in the Moon rather reflect mixing between the proto-Earth and Theia, it would suggest that either Theia was more volatile-depleted than the proto-Earth. Alternatively, Theia was volatile-rich, but the Moon contains a higher fraction of proto-Earth material than the Earth today (i.e., $\delta f_T$ is >0). However, this seems unlikely, given that most of the Giant Impact simulations result in $\delta f_T < 0$. Another interesting observation is the difference in the ratio of the lithophile elements Nb/Ta between Earth and Moon, suggesting that less than 65% (most likely 30-50%) of the Moon is derived from Theia (Münker et al., 2003) if Theia had a chondritic Nb/Ta (like Mars), again compatible with a hit-and-run origin of the Moon. But perhaps most importantly, the FeO contents of the terrestrial mantle and the Moon are different, with a most recent estimate of 7.67% and 10.6%, respectively (Warren & Dauphas, 2014). It is unclear what determines the FeO content of a planetary mantle. At face value, the ~8% FeO in the mantle, and ~80% Fe in the core would suggest that the Earth's equilibrium $f_{O2}$ (oxygen fugacity) is at ~IW-2 (two log units below the iron-wüstite buffer; Wade & Wood, 2005). However, this cannot explain the abundances of Ni and Co in the terrestrial mantle. Wade & Wood (2005) therefore suggest that the Earth's mantle started at a lower $f_{O2}$ (and thus lower FeO) content, and became progressively more oxidized over the course of accretion, perhaps due to perovskite formation in the deep mantle (see also Williams et al., 2012). But such a "self-oxidation" process cannot be the only explanation for the FeO content of a planetary mantle, because Mars or Vesta are not massive enough to stabilize perovskite, yet have even higher mantle FeO values (18% and 26%, respectively; Righter et al., 2006) than the Earth. Therefore, we should expect that the FeO contents of planetary mantles are not a simple function of core-mantle interactions, but can, e.g., also be inherited from the material from which the planets and planetary embryos accreted. Ringwood (1979) and Karato (2014) have suggested that the Moon might have accreted from FeO-enriched, volatile-depleted material derived from the proto-Earth's magma ocean, a suggestion however that can currently not be addressed in SPH simulations of the Giant Impact, which treat planetary mantles as chemically uniform.

If the present-day FeO content of the Earth and Moon is indeed "inherited" from the bulk mantles of the proto-Earth and Theia (for an FeO content <30%, when the cosmic abundance of Fe in Theia would have been fully oxidized), the minimum $|\delta f_T|$ that can still reproduce the observed Earth-Moon difference (2.9%) is about 12% (Figure 10). This rules out all the low $|\delta f_T|$ runs of the high angular momentum models that have been used above to explain the isotopic similarity of the Earth and Moon (Figures 3-7). If a part of the Earth's FeO derives from post-Giant Impact self-oxidation or the addition of a highly oxidized late veneer, the initial Earth-Moon difference (after the Giant Impact) would have been even higher, and with it the required FeO of Theia. Therefore, any post-impact mantle oxidation on Earth will increase the lower limit to $|\delta f_T|$ values even higher than 12%. It seems unlikely that self-oxidation and oxidation by late-veneer material would be more important on the Moon than on Earth. Therefore, if the FeO difference between the Earth and Moon is real (Warren & Dauphas (2014) suggest that an identical FeO value is just barely possible within uncertainties), and inherited from the bulk mantles of the proto-Earth and Theia, the difference in FeO content between the mantles of the Earth and the Moon effectively rules out the high angular momentum models as a viable solution of the isotopic conundrum. Crucially, if the high angular momentum models are excluded, only models with $|\delta f_T| > 12\%$ are "allowed", and Theia must thus have been made from isotopically Earth-like materials (see also table 1). The question whether the FeO content of the Moon is indeed higher than the one of the terrestrial mantle is therefore of high importance for the understanding of both the Giant Impact as well as the question of isotopic heterogeneity of the planet-forming disk.

What about the FeO content of Earth-like materials? NWA 5400 has an FeO content of 25% (Day et al., 2012), suggesting a Theia with a similar FeO content would require a $|\delta f_T|$ of approximately 16%, which should be considered a lower limit if some of the Earth mantle's FeO comes from post-Giant Impact self-oxidation or the addition of a late veneer. On the other hand, the FeO content of enstatite chondrites is close to zero. One might thus argue that an enstatite chondritic Theia is not possible, as it would result in an FeO-depleted Moon relative to Earth. However, a Theia like the one from the hit-and-run model would also be massive enough to incorporate metallic Si into its core (see last sub-section), a process that can contribute to the oxidation of the mantle, as proposed by Javoy (1995):

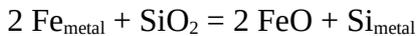

$$2\ Fe_{metal} + SiO_2 = 2\ FeO + Si_{metal}$$

Applying this to the range of 7-20% Si in Theia's core suggested by the isotopic similarity of the Earth and Moon in $\delta^{30}Si$ (Figure 9) under the hit-and-run model, a mantle with an FeO fraction of 10 – 20% results. This is compatible with the required FeO content of Theia to reproduce the Earth-Moon difference under the hit-and-run and canonical Giant Impact models (solid black line in Figure 10). However, under no such Fe/Si self-oxidation scenario can the FeO content of Theia become high enough (i.e., >30%) to be compatible with a $|\delta f_T| < 12\%$, as in the high angular momentum models, as there is simply not enough Fe available within Theia to oxidize.

In summary of this section, if the difference in FeO content between the Earth and Moon is real, and the FeO contents of Earth and Moon are predominantly inherited from the bulk mantles of the proto-Earth and Theia, this puts a lower limit on the $|\delta f_T|$ of the Giant Impact (~12%). This would exclude the recently presented high angular momentum models, and Theia therefore has to be isotopically highly similar to the Earth. The hit-and-run model by Reufer et al. (2012) seems to fit well with an originally enstatite-chondritic Theia, regarding O isotopes (Herwartz et al., 2014), Nb/Ta-ratios (Münker et al., 2003), as long as its mantle was enriched in both isotopically heavy Si and FeO through incorporation of metallic Si (approximately 10%) into the theian core.

*3.7. Summary*

The high angular momentum models (impact fission, merger) allow for Theia to be of CI-chondritic or Mars-like isotopic composition in O, Ti, Cr, Zr, Si and W, as long as $|\delta f_T|$ is smaller than about 4%. However, the very low $|\delta f_T|$ values of the high angular momentum models cannot explain the observed difference between the FeO content of Earth and the Moon, if these FeO contents are predominantly inherited from the proto-Earth and Theia. If Theia had a realistic FeO content <30%, the $|\delta f_T|$ value of the Giant Impact must have been higher than 12%, restricting the isotopic composition of Theia to the class of "Earth-like materials" including enstatite chondrites and achondrites as well as ungrouped achondrites like NWA 5400 and NWA 4741. While these materials are incompatible with the canonical Giant Impact, they are compatible with all three new Giant Impact models. An enstatite chondritic Theia requires some incorporation of Si into its core, and is thus only compatible with Giant Impact models in which Theia is massive enough to start this process (i.e., the hit-and-run and the merger model). An added benefit of this process is that it is capable of elevating the low FeO content of an enstatite chondrite derived mantle to the values required to explain the FeO content of the Moon.

In Table 1, for future reference, we summarize the possible ranges for the isotopic composition of Theia, in the absence of post-impact isotopic re-equilibration, for all four major Giant Impact models.

**4. Conclusions**

We have used the new models of the Giant Impact to provide a possible range of isotopic composition in O, Ti, Cr, Zr, Si and W for Theia, the planetary embryo which collided with the Earth in the Moon-forming impact. We also calculate the likely mantle FeO content of Theia, assuming the Earth and Moon inherited their mantle FeO contents of 7.7% and 10.6% (Warren & Dauphas, 2014), respectively, at least partly from Theia and the proto-Earth. We have shown that the high angular momentum models of the Giant Impact (*impact fission*, by Cúk & Stewart, 2012, and *merger*, by Canup, 2012) can indeed result in the low $|\delta f_T|$ values (<10%) required to resolve the isotopic conundrum, and thus have Theia accreted (theoretically) from any known solar system material in the extreme cases of $|\delta f_T|$ < 1%. However, the implied FeO content of the theian mantle in these models exceeds 30%, the maximum observed in solar system materials. Therefore, Giant Impact models with $|\delta f_T| > 12\%$ are required, e.g., the hit-and-run model (Reufer et al., 2012). If the FeO content of the Earth and the Moon is indeed inherited from the proto-Earth and Theia, then by implication Theia must have had an Earth-like isotopic composition (similar to enstatite chondrites, aubrites or other Earth-like materials like NWA 5400). This is possible if both the Earth and Theia, but not Mars, were part of an early inner disk uniform reservoir (IDUR; Dauphas et al., 2014), or if the inner disk region has been isotopically homogenized in the time between the isolation of Mars from the disk and the Giant Impact that formed the Moon. The existence of such a homogenized inner disk reservoir also requires that Venus has an Earth-like isotopic composition. While a sample-return or in-situ mission to Venus will not be realized in the foreseeable future, systematic searches for, and studies of, Earth-like materials similar to the ungrouped achondrites NWA 5400 and NWA 4741 can and should be pursued to shed more light on the late stages of planet formation.

*Acknowledgments:* The authors thank B. Wood and S.-I. Karato for their helpful reviews, W. M. Akram for discussions, K. Pahlevan and an anonymous reviewer for their comments on an earlier version of this manuscript. This study was supported by grants from the Swiss National Science Foundation (M.M., A.R.) and the Swedish Research Council (M. M.). M.M. also thanks Bernard Marty for his support for this project.

**Figure captions**

**Figure 1:** Giant Impact simulation runs from the literature, from all three new models, plotting the resulting mass of the satellite vs. $|\delta f_T|$. Solid symbols represent simulation runs that were "successful", i.e., that resulted in a satellite with parameters comparable to those of the actual Moon (see main text), while open symbols represent all other ("unsuccessful") simulation runs. Symbol size corresponds to the final angular momentum of the Earth-Moon system after the Giant Impact. While the hit-and-run and impact fission models lead to relatively tight clusters at $|\delta f_T|$ = 35% and 8%, respectively, the merger simulations show considerable scatter, from $|\delta f_T|$= 0% to 40%. Abbreviations in the legend: R12 = Reufer et al., 2012; C12 = Canup, 2012; CS12 = Cúk & Stewart, 2012. The symbol corresponding to the actual Moon is arbitrarily plotted at $|\delta f_T|$= 0% for comparison.

**Figure 2**: The isotopic composition of carbonaceous (closed symbols, blue) and non-carbonaceous (open symbols) material in the solar system, for $\Delta^{17}O$ vs. $\varepsilon^{54}Cr$ (a; left panel) and $\varepsilon^{50}Ti$ vs. $\varepsilon^{54}Cr$ (b; right panel). In both diagrams, the Earth and Moon are plotting on the edge of the non-carbonaceous field, suggesting a higher fraction of carbonaceous material than, e.g., Mars (red). The only primitive (chondritic) materials plotting close the Earth and Moon are enstatite chondrites (green). HED = Howardites, Eucrites, Diogenites. MG Pallasites = Main Group Pallasites. See main text for data sources (after Warren, 2011).

**Figure 3**: The constraints on the O-isotopic composition of Theia. On the ordinate, the "allowed" isotopic composition of Theia in $\Delta^{17}O$ is plotted against $|\delta f_T|$ on the abcissa. Theia's isotopic composition must be within the solid black lines to yield the observed isotopic similarity of Earth and Moon in $\Delta^{17}O$ (without post-impact isotopic equilibration). The more $|\delta f_T|$ approaches ~0%, the wider the range of allowed isotopic composition becomes. The range of $|\delta f_T|$ values in the merger, impact-fission, hit-and-run and canonical Giant Impact models is represented by light gray, gray hatched (top-left-to-down-right; down-left-to-top-right) and black areas, respectively. Down to $|\delta f_T|$ values of about ~20% (i.e., under the canonical and hit-and-run models), enstatite chondrites and other Earth-like materials (light and green, respectively) are the only possible proxies for Theia. For $|\delta f_T|$ values lower than approximately 6%, other known solar system materials (e.g., angrites, Mars, CI-chondrites) enter the allowed range of $\Delta^{17}O$ for Theia. For other carbonaceous chondrites, $|\delta f_T|$

values <1% are required, which are only reached in a few successful runs of the merger model.

**Figure 4:** Contstraints on the $\varepsilon^{50}$Ti composition of Theia as a function of $|\delta f_T|$. Down to $|\delta f_T|$ values of approximately 15%, enstatite chondrites are the only known solar system materials that might be proxies for Theia. CI chondrites are compatible with some simulation runs of the impact-fission model. So far, there are no Ti data for NWA 5400.

**Figure 5:** Constraints on the $\varepsilon^{54}$Cr composition of Theia as a function of $|\delta f_T|$. Although the variation in $\varepsilon^{54}$Cr among solar system materials is similar to, e.g., the variation in $\varepsilon^{50}$Ti, the isotopic composition of Theia is less constrained due to the larger uncertainties of the terrestrial and lunar composition. Earth-like materials (green) are compatible with all models, as is Mars (within uncertainties). Carbonaceous chondrites, and CI chondrites, require low $|\delta f_T|$ values <23%, and <12%, respectively.

**Figure 6:** Constraints on the $\varepsilon^{96}$Zr composition of Theia as a function of $|\delta f_T|$. No data on Mars, angrites or NWA 5400 are available. Enstatite and CI chondrites are compatible with all models. Other types of carbonaceous chondrites are only compatible with low $|\delta f_T|$ <30%, as observed in the merger, impact-fission, and a few simulation runs of the hit-and-run model.

**Figure 7:** Constraints on the $\delta^{30}$Si composition of Theia as a function of $|\delta f_T|$. The Earth and Moon are identical, but set apart from the rest of the known solar system materials in their mass-dependent fractionation of Si isotopes. If Theia's $\delta^{30}$Si composition was chondritic, this requires that the $|\delta f_T|$ of the Moon-forming event was <10%, and even <5% if it was enstatite chondritic. Therefore, the hit-and-run or canonical Giant Impact models require that Theia had fractionated $\delta^{30}$Si as well (see main text). No data are available for NWA 5400.

**Figure 8**: The maximum ranges for the the composition of Theia in $\varepsilon^{182}$W and Hf/W, as a consequence of the similarity of the (pre-late veneer) Earth and Moon in these values. Under the canonical model, Theia must have been almost indistinguishable from the Earth and Moon in these values, but could have been Mars-like under the hit-and-run model. The impact-fission model allows an even broader range of possible values (plotting outside the area). Also shown are the Hf/W and $\varepsilon^{182}$W values of planetary embryos from the Nimmo et al. (2010) chemical accretion model.

**Figure 9:** The isotopic composition of the Earth and Moon in $\delta^{30}$Si as a function of the core Si content of Theia. In both the hit-and-run and merger models, a certain range of core Si fractions are compatible with both the simliarity of Earth and Moon in $\delta^{30}$Si as well as their absolute value in $\delta^{30}$Si. For the hit-and-run model, a core Si content of 7-20% is required, while contents <17% are required in the merger model. For most of the possible core Si contents in the hit-and-run model, the resulting $\delta^{30}$Si is slightly heavier than the one of the Earth. Model parameters are given and explained in the SOM.

**Figure 10:** The FeO content of Theia, as a function of $\delta f_T$, assuming the FeO contents of the Earth and Moon are inherited from the proto-Earth and Theia (black solid line). The impact fission model and the high $\delta f_T$ simulation runs of the merger model clearly cannot account for the FeO difference between the Earth and Moon, as they would require FeO contents >30% (for $|\delta f_T| < 10\%$) to be compatible. The FeO values of Mars, CI chondrites and Mercury are from Righter et al. (2006).

**Figures**

**Figure 1**

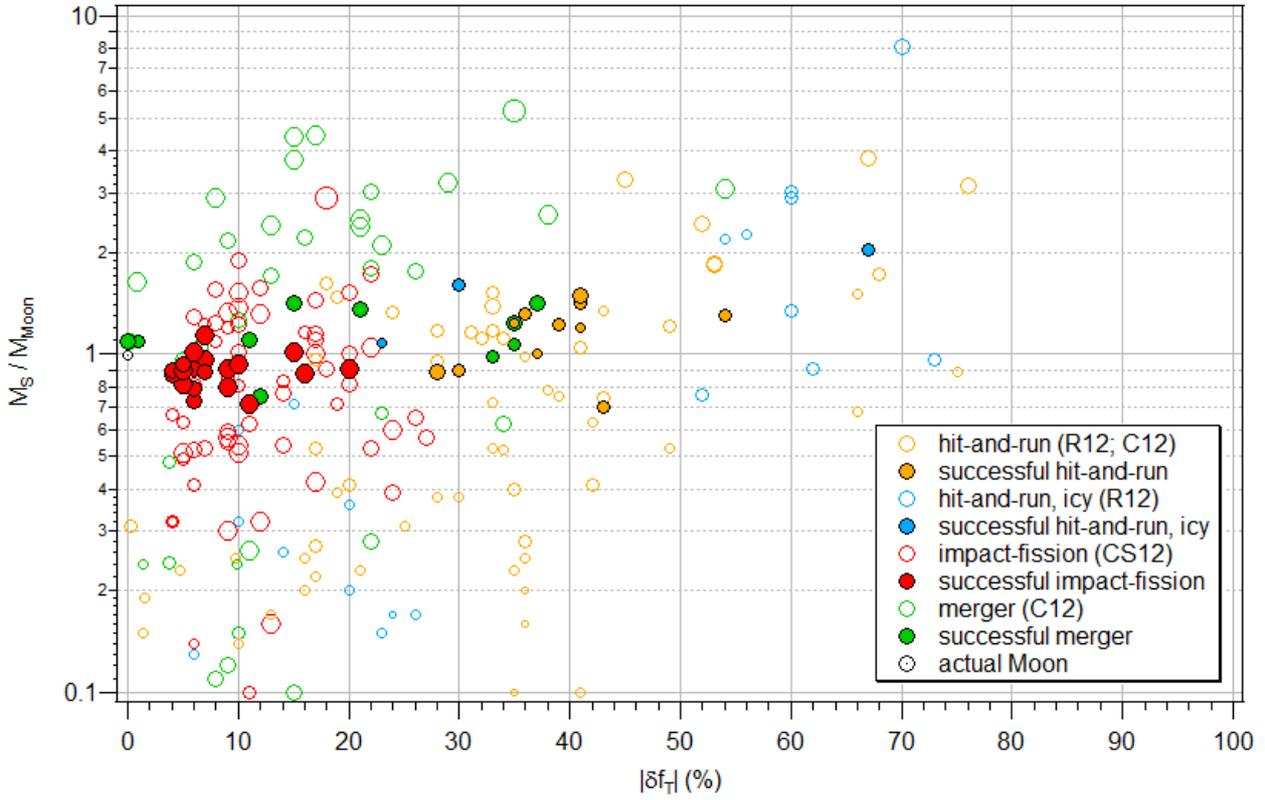

**Figure 2**

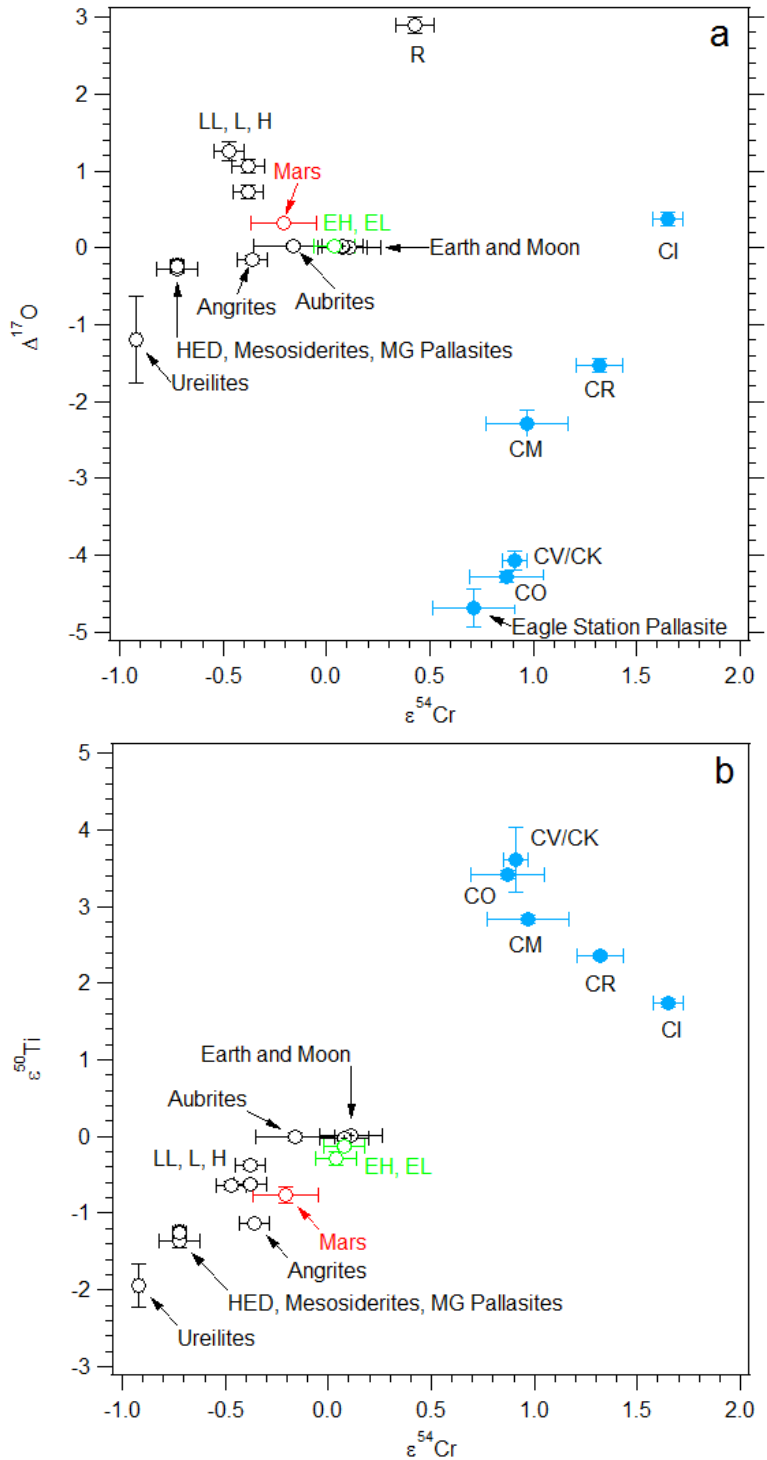

**Figure 3**

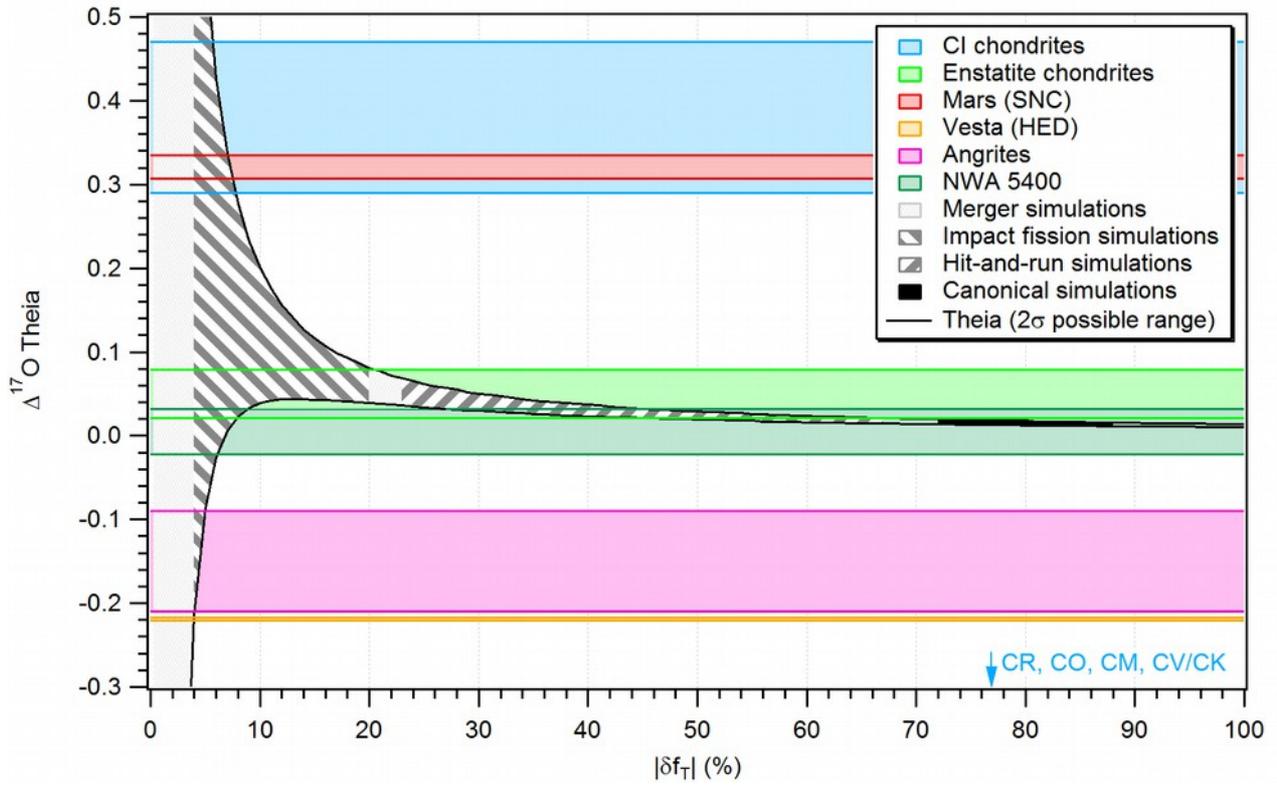

**Figure 4**

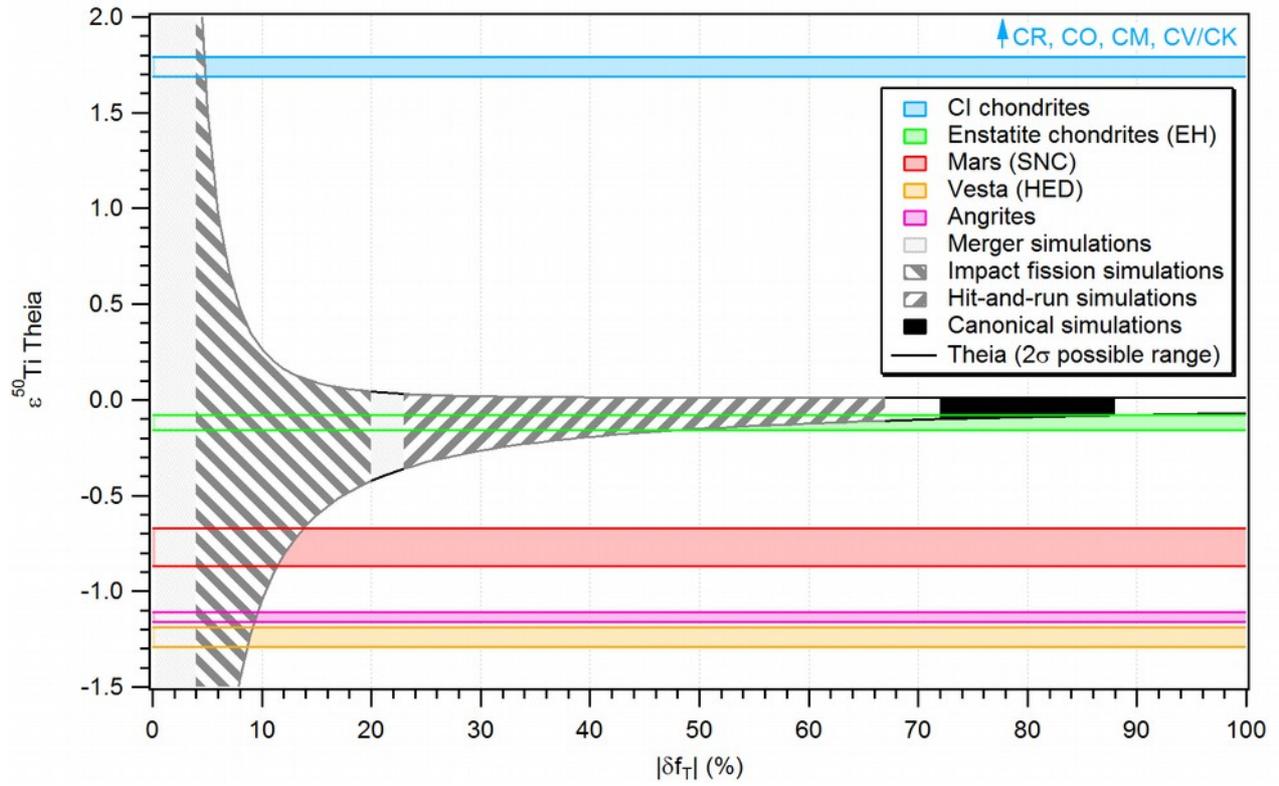

**Figure 5**

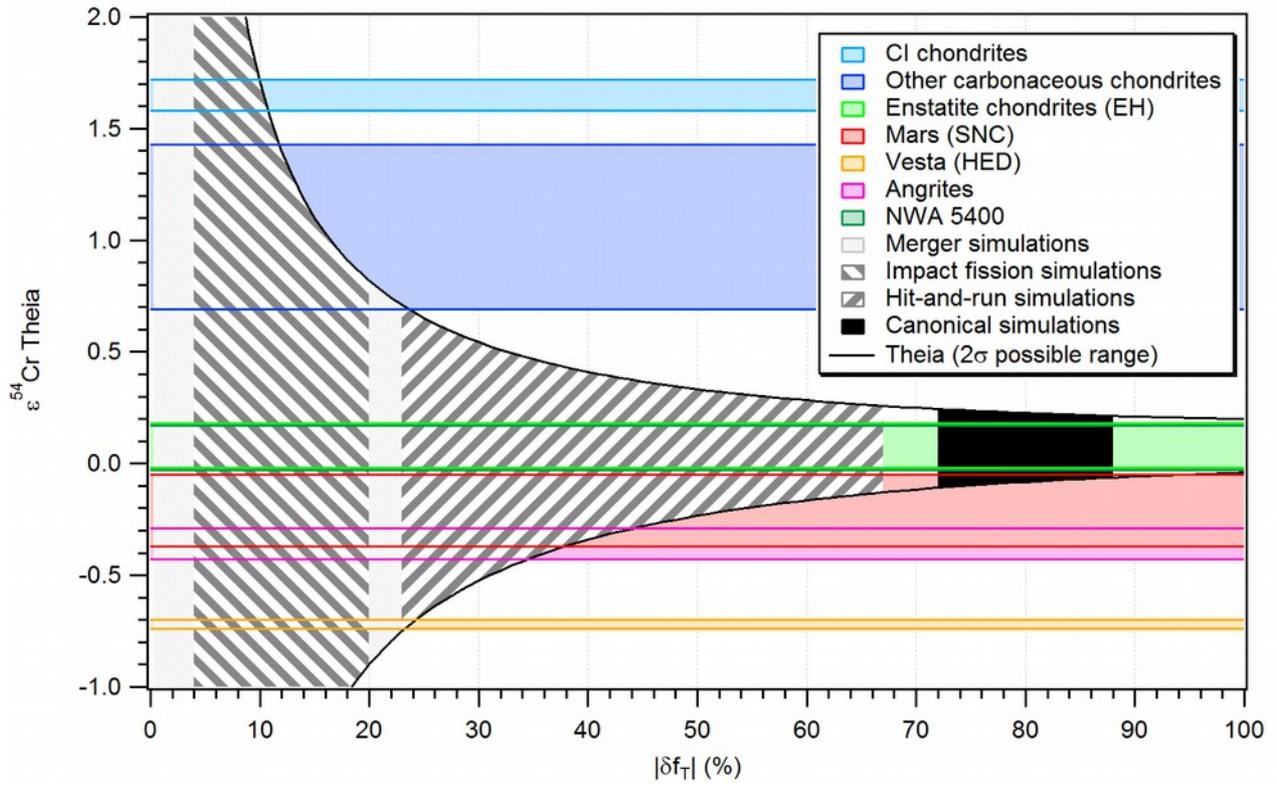

**Fiugre 6**

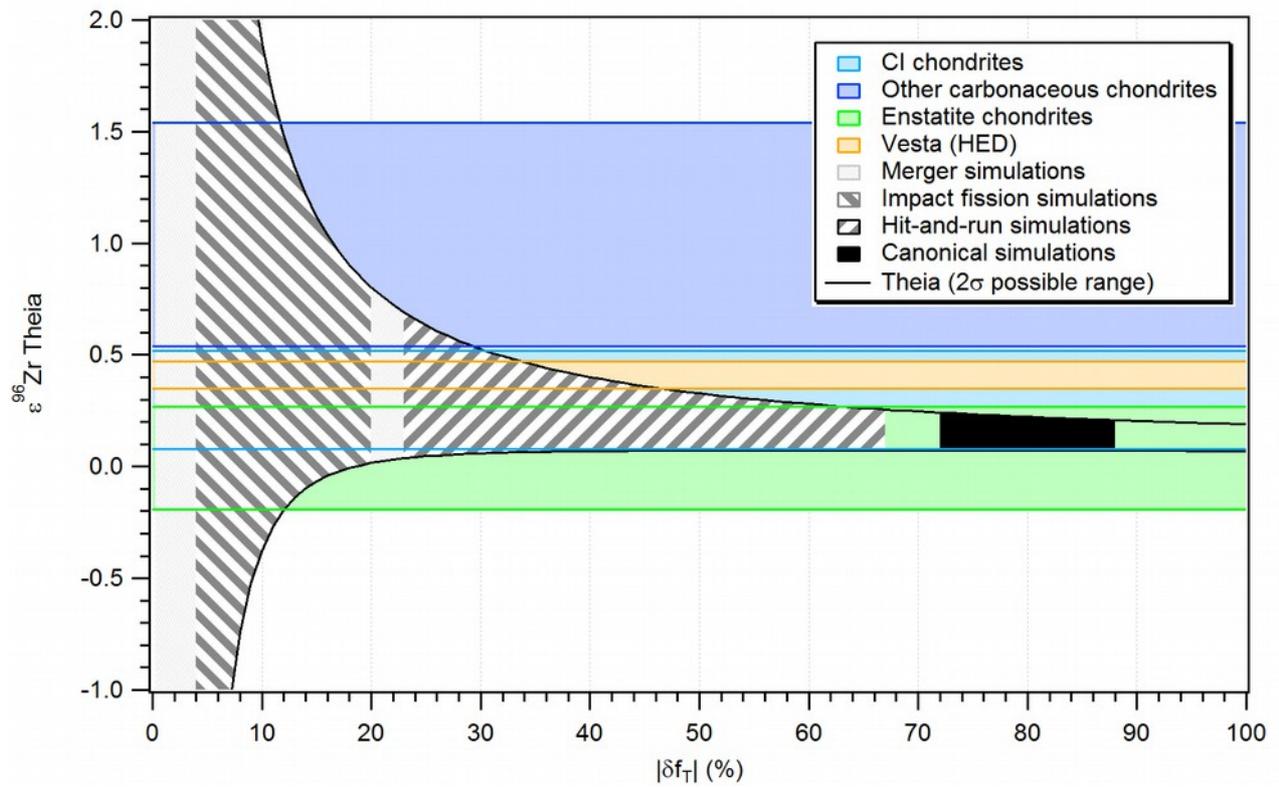

**Fiugre 7**

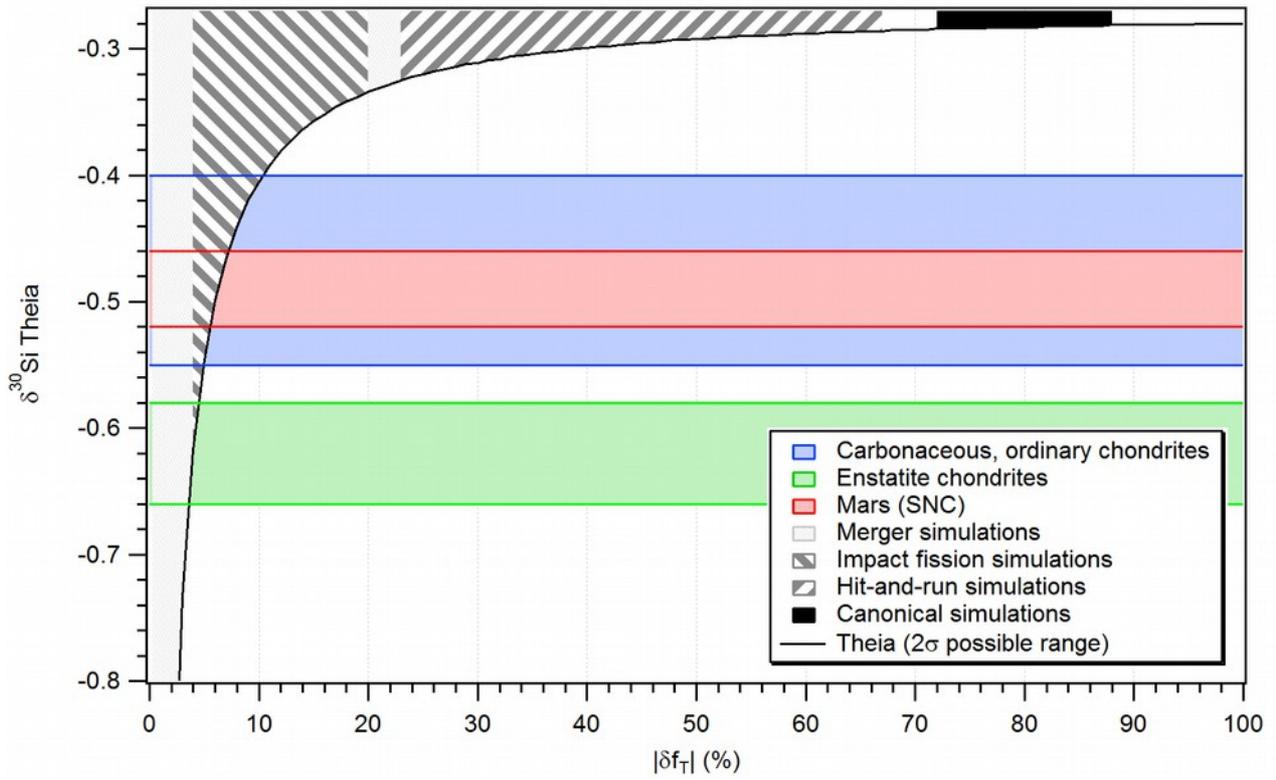

**Figure 8**

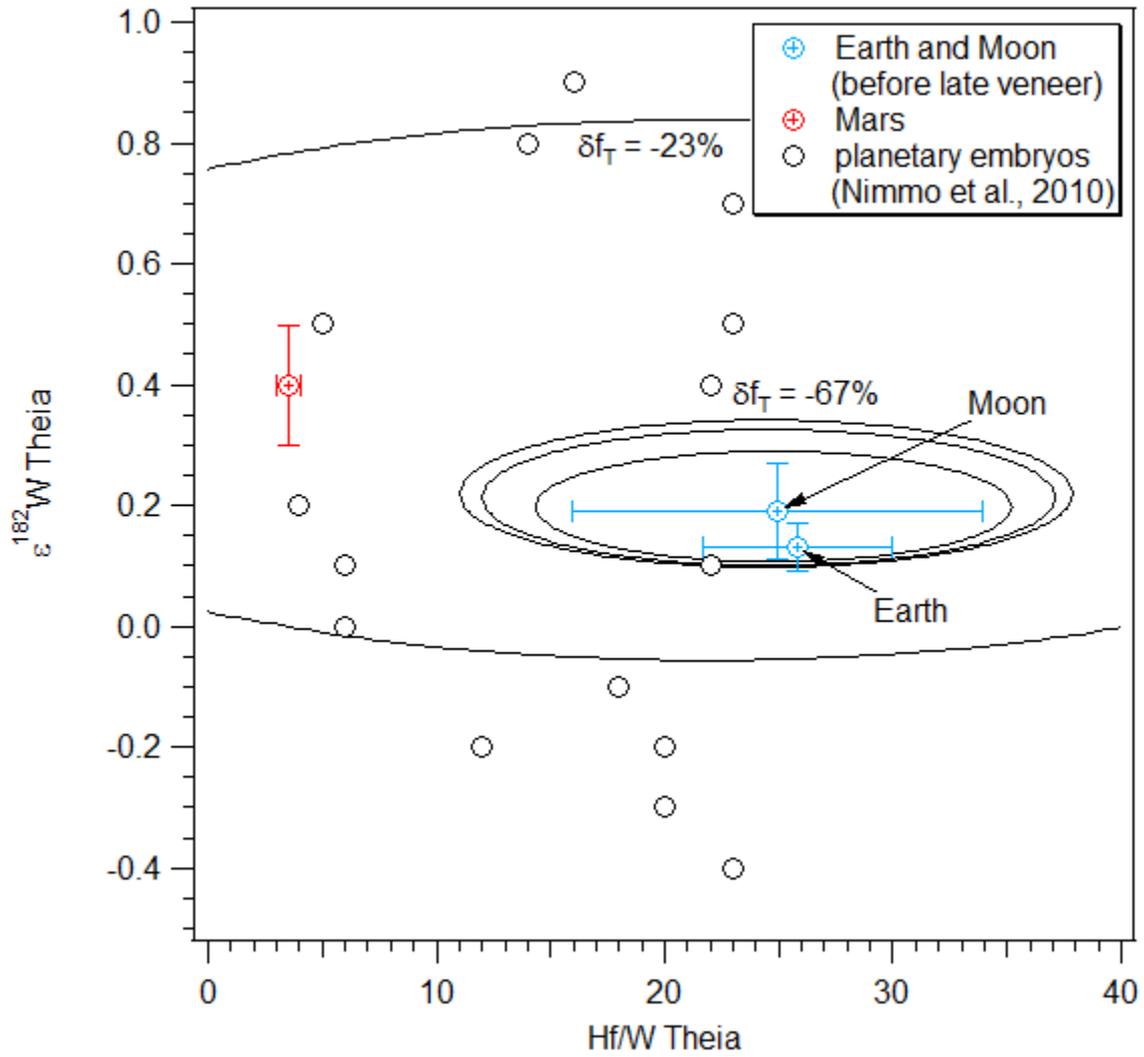

**Figure 9**

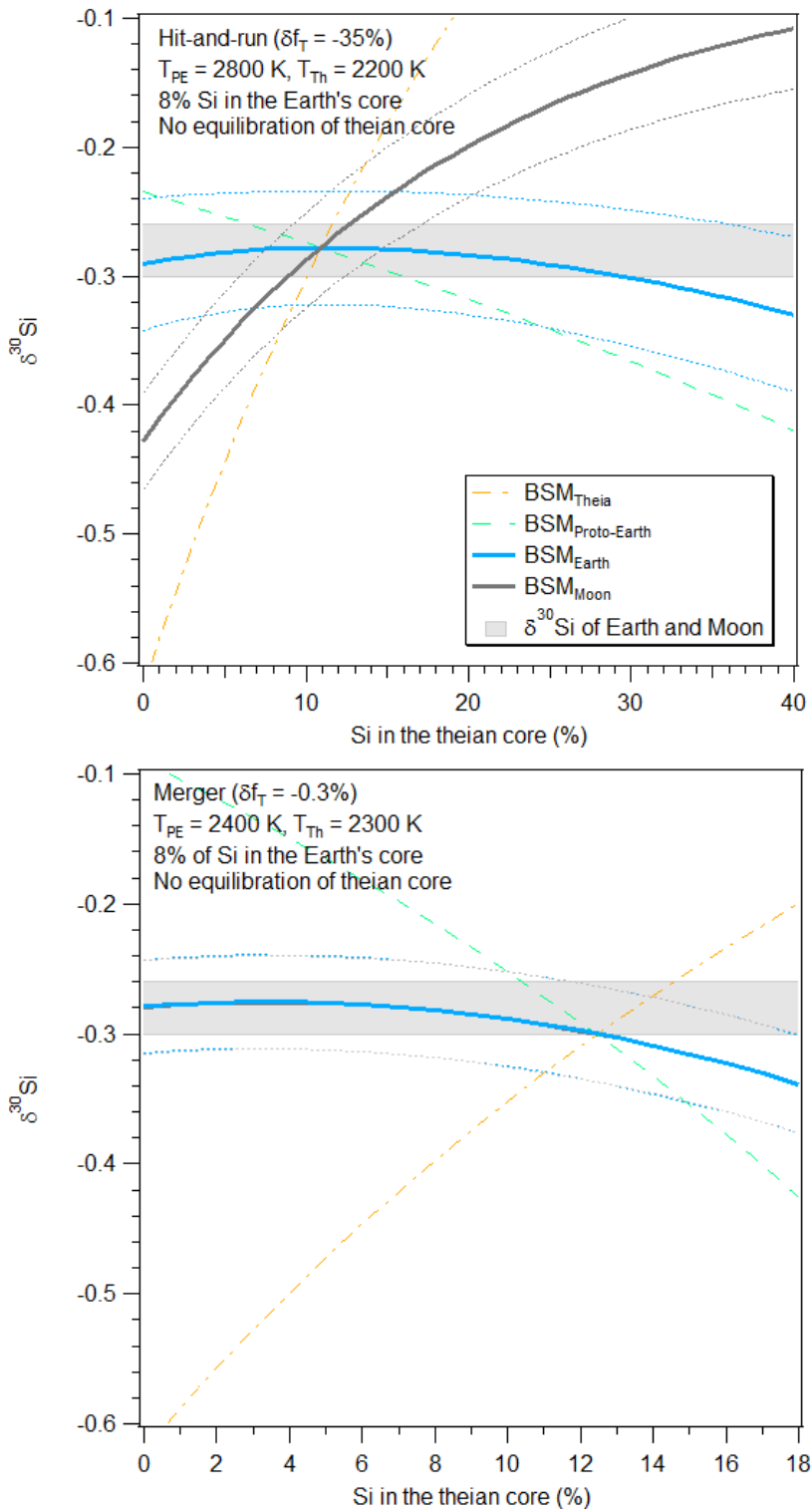

**Figure 10**

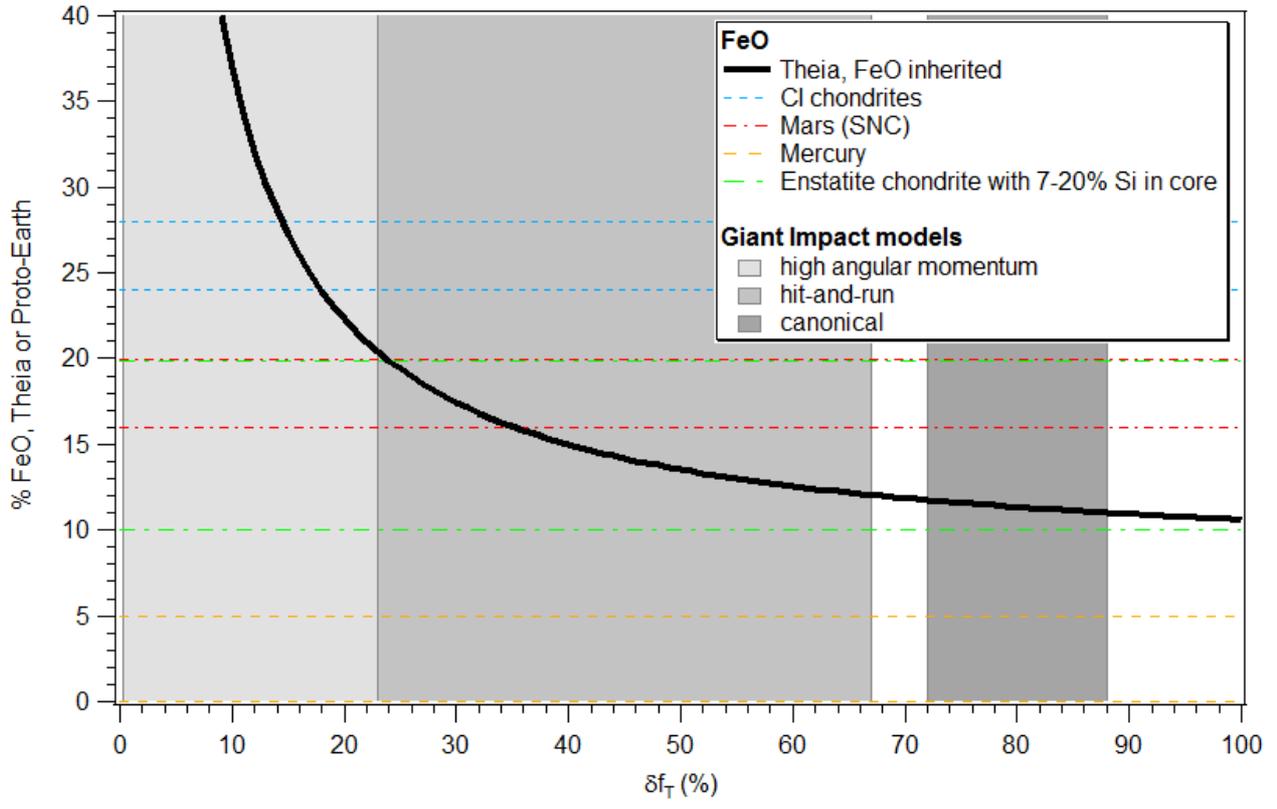

**Tables**

**Table 1: Overview of possible isotopic compositions of Theia**

| Giant Impact Model | $|\delta f_T|$ (%) | $\Delta^{17}O$ | $\varepsilon^{50}Ti$ | $\varepsilon^{54}Cr$ | $\varepsilon^{96}Zr$ | $\delta^{30}Si$ | $\varepsilon^{182}W$* | Hf/W | FeO (%) | Compatible materials** |
|---|---|---|---|---|---|---|---|---|---|---|
| *Low to intermediate angular momentum models* | | | | | | | | | | |
| Canonical (high range) | 88 | +0.016 +0.011 | +0.010 -0.081 | +0.214 -0.062 | +0.208 +0.071 | -0.259 -0.281 | +0.289 +0.107 | +35.2 +14.4 | 11.0 | none |
| Canonical (average) | 80 | +0.018 +0.012 | +0.010 -0.090 | +0.227 -0.082 | +0.223 +0.072 | -0.257 -0.283 | +0.306 +0.104 | +36.0 +13.3 | 11.3 | none |
| Canonical (low range) | 72 | +0.020 +0.014 | +0.010 -0.101 | +0.245 -0.108 | +0.242 +0.072 | -0.256 -0.284 | +0.326 +0.101 | +37.1 +12.0 | 11.7 | none |
| Hit-and-run (high range) | 67 | +0.021 +0.015 | +0.010 -0.110 | +0.259 -0.129 | +0.256 +0.072 | -0.254 -0.286 | +0.341 +0.098 | +37.9 +11.0 | 12.0 | none |
| Hit-and-run (average) | 39 | +0.038 +0.024 | 0.014 -0.199 | +0.421 -0.355 | +0.410 +0.069 | -0.240 -0.300 | +0.509 +0.059 | +46.1 ~0 | 15.2 | ELM, EC*** |
| Hit-and-run (low range) | 23 | +0.069 +0.036 | +0.031 -0.359 | +0.710 -0.750 | +0.691 +0.038 | -0.215 -0.325 | +0.838 -0.056 | +60.0 ~0 | 20.4 | ELM, EC*** |
| *High angular momentum models* | | | | | | | | | | |
| Impact fission (high range) | 20 | +0.081 +0.039 | +0.043 -0.423 | 0.818 -0.898 | +0.803 +0.017 | -0.206 -0.334 | +0.978 -0.118 | +65.1 ~0 | 22.3 | ELM |
| Impact fission (average) | 8.0 | +0.277 +0.023 | +0.471 -1.451 | +2.20 -2.73 | +2.61 -0.738 | -0.100 -0.440 | +0.356 -1.80 | +123 ~0 | (44.3) | ELM |
| Impact fission (low range) | 4.0 | +0.814 -0.214 | +2.56 -4.54 | +5.41 -6.69 | +8.01 -4.39 | +0.077 -0.617 | +12.1 -8.79 | +221 ~0 | (80.9) | CI, Mars, Angr., HED, EC, ELM |
| Merger (high range) | 37 | +0.040 +0.025 | +0.014 -0.211 | +0.443 -0.385 | +0.431 +0.068 | -0.238 -0.302 | +0.532 +0.052 | +47.2 ~0 | 15.6 | ELM, EC*** |
| Merger (average) | 19 | +0.087 +0.040 | +0.049 -0.450 | +0.862 -0.958 | +0.850 +0.006 | -0.202 -0.338 | +1.04 -0.147 | +67.1 ~0 | 23.1 | ELM, EC*** |
| Merger (low range) | 0.3 | +98.0 -90.0 | +613 -640 | +698 -718 | +1120 -1070 | +4.44 -4.98 | +1900 -1860 | +2600 ~0 | (>100) | all known |

*The two numbers given in each cell correspond to the upper and lower end of the 2σ confidence interval for the isotopic composition of Theia, for each of the Giant Impact scenarios and elements. ELM = Earth-like materials (see main text); EC = enstatite chondrites; CI = carbonaceous chondrites of type CI. *Based on pre-late veneer values of Earth and Moon. **Without taking the FeO content into account. ***After accounting for Si incorporation into Theia's core and corresponding Si isotope fractionation.*

**Supplementary Material for "On the origin and composition of Theia: Constraints from new models of the Giant Impact" by Meier et al., 2014.**

**Overview / table of contents**

*1. Units of isotopic ratio deviations*
*2. Definition of parameter $\delta f_T$*
*3. Mass balance calculations*
*Supplementary References*

*1. Units of isotopic ratio deviations*

Deviation in the isotpic ratio of isotope $^iX$ and reference isotope $^jX$ of element X:

In permil: $\delta^i X = [\ (^iX/^jX)_{sample} /\ (^iX/^jX)_{standard} - 1\ ] \times 1000$ ‰

In parts per $10^4$: $\varepsilon^i X = [\ (^iX/^jX)_{sample} /\ (^iX/^jX)_{standard} - 1\ ] \times 10000 \varepsilon$

Mass-independent sfractionation of $^{17}O$: $\Delta^{17}O = \delta^{17}O - 0.52 \times \delta^{18}O$

**2. *Definition of parameter $\delta f_T$***

The parameter $\delta f_T$ is a measure for the difference of the fraction of material from the "target" (the proto-Earth) in the Moon-forming disk and the Earth, respectively:

In percent: $\delta f_T = [\ (M_{Si,targ} / M_{Si,tot})_{Disk} / (M_{Si,targ} / M_{Si,tot})_{Earth} - 1\ ] \times 100\%$

As an example, let us consider the canonical model, where Theia contributes 80% to the silicate mass of the Moon-forming disk, and $0.1 \times 0.68$ Earth masses to the silicate Earth (assuming Earth and Theia had identical core mass fractions of 32%). In this case, the proto-Earth contributed 0.2 Moon masses of silicates to the Moon, and $0.61 = 0.68 - 0.068$ Earth masses of silicates to the Earth:

$[(0.20/0.95) / (0.61/0.68) – 1] \times 100\% = -77\%$.

The parameter $\delta f_T$ is preferred here to other possible notations because it allows us to express isotopic similarity of the Earth and Moon without having to care about the relative contributions of Theia to the Earth and Moon in the different Giant Impact models.

**3.** *Mass balance calculations*

*3.1. The isotopic composition of Theia as a function of $\delta f_T$ and the isotopic compositions of the Moon and the Earth.*

The ratio of two isotopes of an element in the Earth today ($X_E$) is a function of the respective ratio in Theia ($X_{Th}$), the proto-Earth ($X_{PE}$), the relative mass contribution of Theia to the Earth ($f_{T(E)}$), as well as the relative concentrations of the respective element in Theia ($c_{Th}$) and the proto-Earth ($c_{PE}$):

$$X_E = [\, f_{T(E)} \times c_{PE} \times X_{PE} + (1 – f_{T(E)}) \times c_{Th} \times X_{Th} \,] / [\, f_{T(E)} \times c_{PE} + (1 – f_{T(E)}) \times c_{Th} \,] \quad (1)$$

In analogy, the isotopic composition of the Moon today can be expressed as:

$$X_M = [\, f_{T(M)} \times c_{PE} \times X_{PE} + (1 – f_{T(M)}) \times c_{Th} \times X_{Th} \,] / [\, f_{T(M)} \times c_{PE} + (1 – f_{T(M)}) \times c_{Th} \,] \quad (2)$$

The denominators in equations 1 and 2 correspond to the present-day concentrations of the repsective element in the bulk Earth ($c_E$) and Moon ($c_M$), respectively:

$$f_{T(E)} \times c_{PE} + (1 – f_{T(E)}) \times c_{Th} = c_E \quad (3)$$

$$f_{T(M)} \times c_{PE} + (1 – f_{T(M)}) \times c_{Th} = c_M \quad (4)$$

We then re-arrange equation (1) to get $X_{PE}$:

$$X_{PE} = [\, c_E \times X_E – (1 – f_{T(E)}) \times c_{Th} \times X_{Th} \,] / [\, f_{T(E)} \times c_{PE} \,] \quad (5)$$

We use (5) to replace $X_{PE}$ in equation (2) to get a mixing equation for the isotopic composition of the Moon, that does not include the (unknown) isotopic composition of the proto-Earth anymore:

$$X_M = [\,(f_{T(M)} / f_{T(E)}) \times (c_E \times X_E - (1 - f_{T(E)}) \times c_{Th} \times X_{Th}) + (1 - f_{T(M)}) \times c_{Th} \times X_{Th}\,] / c_M \qquad (6)$$

or:

$$X_M = [\,(f_{T(M)} / f_{T(E)}) \times (c_E \times X_E) + (1 - (f_{T(M)} / f_{T(E)})) \times c_{Th} \times X_{Th})\,] / c_M \qquad (7)$$

We replace the fractional contributions of the proto-Earth to the Moon ($f_{T(M)}$) and the Earth ($f_{T(E)}$) by the parameter $\delta f_T$:

$$\delta f_T = (f_{T(M)} / f_{T(E)}) - 1 \qquad (8)$$

We re-arrange (7) to get an expression for the isotopic composition of Theia:

$$X_{Th} = [\,(\delta f_T + 1) \times (c_E \times X_E) - c_M \times X_M\,] / [\,\delta f_T \times c_{Th}\,] \qquad (9)$$

Using (3) and (4), we derive that:

$$c_{Th} = [\,(\delta f_T + 1) \times c_E - c_M\,] / \delta f_T \qquad (10)$$

We use (10) to replace $c_{Th}$ in equation (9) and get our final mixing equation:

$$X_{Th} = [\,(\delta f_T + 1) \times (c_E \times X_E) - c_M \times X_M\,] / [\,(\delta f_T + 1) \times c_E - c_M\,] \qquad (11)$$

This equation expresses the isotopic composition of Theia as a simple function of $\delta f_T$, the isotopic composition and the relative mass fractions of the respective element in the (silicate part of the) Earth and the Moon.

Using the known (2σ) uncertainties in isotopic compositions $\Delta X_E$ and $\Delta X_M$, and the adopted uncer-tainties in the mass fraction of the respective element $\Delta c_E$ and $\Delta c_M$ for Earth and Moon, respectively, the propagated uncertainty of $X_{Th}$ is given by the square root of the sum of the squares of the partial derivatives of equation (11), multiplied with the individual uncertainties:

$$\Delta X_{Th} = [ (\{(\delta f_T + 1) \times c_E / ((\delta f_T + 1) \times c_E - c_M)\} \times \Delta X_E)^2 + (\{-c_M / ((\delta f_T + 1) \times c_E - c_M)\} \times \Delta X_M)^2 + (\{(\delta f_T + 1) \times c_M \times (X_M - X_E) / ((\delta f_T + 1) \times c_E - c_M)^2\} \times \Delta c_E)^2 + (\{(\delta f_T + 1) \times c_E \times (X_E - X_M) / ((\delta f_T + 1) \times c_E - c_M)^2\} \times \Delta c_M)^2 ]^{0.5} \qquad (12)$$

Equations (11) and (12) were then used to construct the upper ($X_{Th} + \Delta X_{Th}$) and lower ($X_{Th} - \Delta X_{Th}$) bounds on the isotopic composition of Theia, shown in Figures 3 – 7.

### *3.2. The isotopic composition of the Earth and the Moon under the hit-and-run and merger models if Theia has a bulk Si composition like enstatite chondrites*

For Figure 8, we start with the equation that yields the isotopic fractionation of Si between silicates and metal (taken from Zambardi et al., 2013; see references therein):

$$\varepsilon_{Si}(T) = \delta^{30}Si_{silicates} - \delta^{30}Si_{metal} = B \times 10^6 / T^2 \qquad (13)$$

B is a number that has been determined empirically by Ziegler et al. (2010) to 7.64±0.47. T is the temperature (in K) at which the silicate-metal separation takes place.

Model work (e.g., Georg et al., 2007; Fitoussi et al., 2009; Fitoussi & Bourdon, 2012) has set this fractionation of Si in a linear relationship to the amount of Si incorporated into the core (at constant temperature):

$$\varepsilon_{Si}(T) = \Delta^{30}Si_{BSE-bulk} / F_{Core} \qquad (14)$$

Here, $F_{Core}$ is the fraction of the Earth's Si that is in the core, and $\Delta^{30}Si_{BSE-bulk}$ is the observed difference in $\delta^{30}Si$ between the bulk silicate Earth and its suspected bulk $\delta^{30}Si$.

The concentration of Si in the core is then:

$$C_{Si,\ core} = (m_{BSE} / m_{Core}) \times [\ F_{Core} / (1 - F_{Core})\ ] \times C_{Si,\ BSE} \qquad (15)$$

Here, $m_{BSE} / m_{Core}$ is the mass ratio of the bulk silicate Earth vs. the core (2.1) and $C_{Si,\ BSE}$ is the concentration of Si in the bulk silicate Earth (21%; McDonough & Sun, 1995).

We now combine equations 13, 14 and 15 and re-arrange to get $\Delta^{30}Si_{BSE\text{-}bulk}$ as a function of $C_{Si,\ core}$:

$$\Delta^{30}Si_{BSE\text{-}bulk} = (B \times 10^6 / T^2) \times [\ 2.27 \times C_{Si,\ core} / (1 + 2.27 \times C_{Si,\ core})\ ] \qquad (16)$$

We add $\Delta^{30}Si_{BSE\text{-}bulk}$ to the assumed bulk $\delta^{30}Si$ of an enstatite chondritic Theia (-0.62±0.05) to get the predicted $\delta^{30}Si$ of the bulk silicate mantle of Theia as a function of the Si core mass fraction. The derivation of the typical temperature of Si-metal equilibration is explained below. This equation was used to construct the graph of Theia in Figure 9 (in orange).

If we assume the temperature to be precisely known, the only error that needs propagation is the one of B. Therefore, the error of the value above is simply:

$$\Delta\Delta^{30}Si_{BSE\text{-}bulk} = (10^6 / T^2) \times [\ 2.27 \times C_{Si,\ core} / (1 + 2.27 \times C_{Si,\ core})\ ] \times \Delta B \qquad (17)$$

The fraction of the Si in the Earth's core that was contributed by Theia is a function of the mass contributed by the theian core ($M_{Th\text{-}core}$) and the present-day Si concentration of the core ($C_{Si,\ E\text{-}core}$). The Si content of the proto-Earth's core ($C_{Si,\ PE\text{-}core}$) was calculated as follows:

$$C_{Si,\ PE\text{-}core} = (C_{Si,\ E\text{-}core} - M_{Th\text{-}core}) / (1 - M_{Th\text{-}core}) \qquad (18)$$

Then, the $\delta^{30}Si$ of the bulk silicate mantle of the proto-Earth (and the uncertainty of that value) was calculated with equation 16 (assuming an ordinary/carbonaceous chondritic bulk $\delta^{30}Si$).
We modify the mass balance calculations developped in 3.1. to calculate the isotopic composition of the Earth and Moon based on the $\delta^{30}Si$ values of the bulk silicate mantles of Theia and the proto-Earth. We start with equations 1 and 2 (here, X stands for $\delta^{30}Si$):

$X_E = [\ f_{T(E)} \times c_{PE} \times X_{PE} + (1 - f_{T(E)}) \times c_{Th} \times X_{Th}\ ] / [\ f_{T(E)} \times c_{PE} + (1 - f_{T(E)}) \times c_{Th}\ ]$ (1)

$X_M = [\ f_{T(M)} \times c_{PE} \times X_{PE} + (1 - f_{T(M)}) \times c_{Th} \times X_{Th}\ ] / [\ f_{T(M)} \times c_{PE} + (1 - f_{T(M)}) \times c_{Th}\ ]$ (2)

We first calculate the mixing lines for the hit-and-run model by Reufer et al. (2012). The average $f_{T(M)}$ from the successful simulation runs is 53±3%. Combined with the average $\delta f_T$ value of the hit-and-run model, -39%, this yields an average $f_{T(E)}$ of 87±6%. We assume that the bulk silicate mantle of Theia has the Si concentration of enstatite chondrites (EH), i.e., 23.5% (Wasson & Kallemeyn, 1988). We can then calculate the required Si concentration in the bulk silicate mantle of the proto-Earth to arrive at today's 21% after re-arranging equation 3:

$c_{PE} = [\ c_E - (1 - f_{T(E)}) \times c_{Th}\ ] / f_{T(E)}$ (19)

This results in a $c_{PE}$ value of 20.6%. We then used this values (and their propagated errors) to calculate the graphs representing the Earth and the Moon (blue and grey, respectively) in Figure 9a.

For the merger model by Canup (2012), we selected the model with the lowest $|\delta f_T|$ of 0.3%. In this simulation run, Nr. 17 from table S5 in Canup (2012), Theia has 45% of the mass of the proto-Earth, the total mass of the system is 1.038 $M_E$, and 0.038 $M_E$ escape after the collision. While the exact values are not given, we assume that the proto-Earth contributes 54.9% of the mass of the Earth, and 55.1% of the mass of the Moon, which results in a $\delta f_T$ = -0.3%. These values were then used to calculate the graphs in Figure 9b.

The fractionation effect on Si isotopes depends on pressure and temperature. Therefore, we had to reconstruct the pressure and temperature conditions for the four planetary objects involved in our mixing calculations (hit-and-run proto-Earth, Theia and merger proto-Earth, Theia). As has been done before, e.g., by Wade & Wood (2005), we assumed that metal-Si equilibration takes place at the peridotite liquidus, at the base of the magma oceans of the planetary objects involved. The base of the magma ocean was set at 35% of the depth of the core mantle boundary, as in Wood et al. (2008). To know this depth, we first calculated the planetary radius and the depth of the core mantle boundary. The former was done using the scaling laws by Seager et al. (2007), while the second was done assuming a constant core mass fraction and a core density that increases linearly from a Mars-like value of 8 g/cm$^3$ at 0.1 M$_E$ to an Earth-like value of 11 g/cm$^3$ at 1 M$_E$. Pressure was calculated assuming a constant gravitational acceleration within the mantle. The temperature at the peridotite liquidus was calculated using the parametrization of Wade & Wood (2005). This temperature increases through time. Since lower temperatures result in stronger Si isotopic fractionation, fractionation is strongest when the planetary object is still small. Si that has once become incorporated into the core is not re-extracted again into the magma ocean (Wade & Wood, 2005). Therefore, the temperature at the time of collision is not necessarily typical of the temperature throughout accretion. We therefore grew our planetary objects in 0.01 M$_E$ intervals and thereby calculated the average temperature of Si equilibration during accretion. All model values are given in supplementary table S3.

**Supplementary References**

References mentioned in the supplementary material and not already mentioned in the main text.

**Table S1: Isotopic composition of solar system materials used in the mixing calculations**

| Material | $\Delta^{17}O$ | $\varepsilon^{50}Ti$ | $\varepsilon^{54}Cr$ | $\varepsilon^{96}Zr$ | $\delta^{30}Si$ | $\varepsilon^{182}W$ | Sources |
|---|---|---|---|---|---|---|---|
| Earth | 0.000±0.002* | 0.01±0.01 | 0.11±0.15 | 0.06±0.04 | -0.27±0.01 | 0.13±0.04 | 1,2,3,4,5,6 |
| Moon | 0.012±0.003* | -0.03±0.04 | 0.08±0.12 | 0.13±0.06 | -0.27±0.01 | 0.19±0.08 | 1,2,3,4,5,7 |
| *Carbonaceous chondrites* | | | | | | | |
| CI | 0.38±0.09 | 1.74±0.05 | 1.65±0.07 | 0.30±0.22 | -0.44±0.03 | -2.2±0.4 | 8,2,3,4,5,10 |
| CM | -2.29±0.18** | 2.84±0.05 | 0.97±0.20 | 0.72±0.14 | -0.52±0.03 | -1.9±0.4 | 8,2,3,4,9,10 |
| CO | -4.28±0.07** | 3.42±0.06 | 0.87±0.18 | 0.72±0.18 | -0.45±0.04 | -2.0±0.4 | 8,2,3,4,5,10 |
| CR | -1.53±0.09** | 2.35±0.04 | 1.32±0.11 | 1.16±0.38 | - | -1.9±0.4 | 8,2,3,4,10 |
| CV/CK | -4.07±0.13** | 3.61±0.43 | 0.91±0.06 | 1.08±0.11 | -0.41±0.01 | -2.1±0.4 | 8,2,3,4,5,10 |
| *Ordinary and R chondrites* | | | | | | | |
| H | 0.73±0.09 | -0.37±0.05 | -0.38±0.07 | | | -2.4±0.5 | 11,2,12,4,5,10 |
| L | 1.07±0.09 | -0.62±0.04 | -0.38±0.08 | 0.41±0.12 | -0.46±0.02 | -2.1±0.5 | 11,2,12,4,5,10 |
| LL | 1.26±0.12 | -0.64±0.06 | -0.47±0.07 | | | -1.5±0.5 | 11,2,12,4,5,10 |
| R | 2.9±0.1 | - | 0.43±0.09 | . | - | - | 13,3 |
| *Enstatite chondrites* | | | | | | | |
| EH | 0.035±0.009* | -0.12±0.04 | 0.08±0.10 | 0.04±0.23 | -0.62±0.04 | -2.3±0.4 | 1,2,3,4,5,10 |
| EL | 0.060±0.007* | -0.29±0.09 | 0.04±0.10 | | | | 1,2,3,4,5,10 |
| *Achondrites* | | | | | | | |
| SNC (Mars) | 0.321±0.014 | -0.77±0.10 | -0.21±0.16 | - | -0.49±0.03 | ~0.4±0.1 | 14,15,3,5,10 |
| HED (Vesta) | -0.219±0.002 | -1.24±0.05 | -0.72±0.02 | 0.41±0.06 | -0.41±0.03 | 22±8 | 16,2,12,4,5,10 |
| Angrites | -0.15±0.06 | -1.14±0.02 | -0.36±0.07 | - | - | - | 17,2,12 |
| Aubrites | 0.02±0.04 | 0.00±0.05 | -0.16±0.19 | - | -0.53±0.04 | - | 17,2,12,9 |
| Ureilites | -1.20±0.57 | -1.95±0.28 | -0.92±0.02 | - | -0.47±0.12 | - | 17,15,18,9 |
| NWA5400 | 0.005±0.027 | - | 0.07±0.10 | - | - | - | 19,19 |
| *Stony Irons* | | | | | | | |
| MG Pallasites | -0.28±0.06 | -1.37±0.08 | -0.72±0.10 | - | - | - | 17,15,12 |
| ES Pallasites | -4.68±0.25 | - | 0.72±0.20 | - | - | - | 17,20 |
| Mesosiderites | -0.24±0.09 | -1.25±0.07 | -0.72±0.03 | - | - | - | 17,15,12 |

*\*To enforce compatibility with earlier measurements, values by Herwartz et al. (2014) were renormalized so that the Earth's mantle is at $\Delta^{17}O = 0.000$. \*\*Not explicitly given in source, calculated as average from all given values of that class. All errors are 2σ, unless stated otherwise. Sources: [1] Herwartz et al., 2014; [2] Zhang et al., 2012; [3] Qin et al., 2010; [4] Akram, 2013; [5] Zambardi et al., 2013; [6] Willbold et al., 2011; [7] Kleine et al., 2014 and Touboul et al., 2014; [8] Clayton & Mayeda, 1999; [9] Armytage et al., 2011; [10] Kleine et al., 2009; [11] Clayton et al., 1991; [12] Trinquier et al., 2007; [13] Kallemeyn et al., 1996; [14] Franchi et al., 1999; [15] Trinquier et al., 2009; [16] Wiechert et al., 2004; [17] Clayton & Mayeda, 1996; [18] Yamakawa et al., 2010; [19] Shukolyukov et al., 2010; [20] Shukolyukov & Lugmair, 2006;*

**Table S2: successful Giant Impact simulation runs from the literature**

| Type | $M_{Theia}$ | $M_{Moon}$ | $Fe_{Disk}$ | $L_{Final}$ | $\delta f_T$ (%) | Source |
|---|---|---|---|---|---|---|
| hit-and-run | 0.15 | 1.20 | 0.03 | 1.16 | -41 | R12 |
| hit-and-run | 0.20 | 0.90 | 0.02 | 1.40 | -30 | R12 |
| hit-and-run | 0.20 | 1.01 | 0.04 | 1.27 | -37 | R12 |
| hit-and-run | 0.20 | 1.32 | 0.02 | 1.46 | -36 | R12 |
| hit-and-run | 0.20 | 1.24 | 0.01 | 1.28 | -35 | R12 |
| hit-and-run | 0.20 | 1.30 | 0.04 | 1.74 | -54 | R12 |
| hit-and-run | 0.20 | 1.08 | 0.00 | 1.09 | -23 | R12 |
| hit-and-run | 0.20 | 1.61 | 0.00 | 1.42 | -30 | R12 |
| hit-and-run | 0.20 | 2.04 | 0.01 | 1.71 | -67 | R12 |
| hit-and-run | 0.20 | 1.22 | 0.03 | 1.41 | -39 | C12 |
| hit-and-run | 0.20 | 1.41 | 0.04 | 1.43 | -41 | C12 |
| hit-and-run | 0.30 | 0.70 | 0.02 | 1.73 | -43 | C12 |
| hit-and-run | 0.30 | 0.89 | 0.00 | 1.88 | -28 | C12 |
| hit-and-run | 0.30 | 1.49 | 0.02 | 1.99 | -41 | C12 |
| merger | 0.40 | 1.10 | 0.02 | 2.18 | 11 | C12 |
| merger | 0.40 | 1.41 | 0.05 | 2.39 | -15 | C12 |
| merger | 0.40 | 1.07 | 0.03 | 1.84 | -35 | C12 |
| merger | 0.40 | 1.24 | 0.04 | 2.24 | -35 | C12 |
| merger | 0.40 | 0.98 | 0.02 | 1.71 | -33 | C12 |
| merger | 0.45 | 1.09 | 0.03 | 1.77 | -1 | C12 |
| merger | 0.45 | 0.75 | 0.02 | 1.94 | -12 | C12 |
| merger | 0.45 | 1.09 | 0.03 | 2.22 | 0.3 | C12 |
| merger | 0.45 | 1.36 | 0.01 | 2.00 | 21 | C12 |
| merger | 0.45 | 1.36 | 0.01 | 2.00 | 21 | C12 |
| merger | 0.45 | 1.42 | 0.00 | 2.03 | 37 | C12 |
| merger | 0.45 | 0.91 | 0.01 | 2.24 | 9 | C12 |
| impact fission | 0.03 | 0.71 | 0.02 | 2.50 | -11 | CS12 |
| impact fission | 0.05 | 0.73 | 0.04 | 2.33 | -6 | CS12 |
| impact fission | 0.10 | 0.80 | 0.04 | 2.49 | -9 | CS12 |
| impact fission | 0.05 | 0.79 | 0.05 | 1.97 | -6 | CS12 |
| impact fission | 0.03 | 0.82 | 0.00 | 2.43 | -5 | CS12 |
| impact fission | 0.03 | 0.87 | 0.01 | 2.32 | -4 | CS12 |
| impact fission | 0.03 | 0.90 | 0.05 | 2.22 | -4 | CS12 |
| impact fission | 0.03 | 0.88 | 0.04 | 2.76 | -16 | CS12 |
| impact fission | 0.05 | 0.91 | 0.00 | 2.56 | -9 | CS12 |
| impact fission | 0.05 | 0.96 | 0.01 | 2.40 | -7 | CS12 |
| impact fission | 0.05 | 0.91 | 0.03 | 2.22 | -6 | CS12 |
| impact fission | 0.05 | 0.93 | 0.01 | 2.67 | -10 | CS12 |
| impact fission | 0.05 | 1.14 | 0.02 | 2.52 | -7 | CS12 |
| impact fission | 0.05 | 1.02 | 0.02 | 2.84 | -15 | CS12 |
| impact fission | 0.10 | 0.89 | 0.02 | 2.30 | -7 | CS12 |
| impact fission | 0.10 | 0.91 | 0.01 | 2.90 | -20 | CS12 |
| impact fission | 0.05 | 1.02 | 0.01 | 2.47 | -6 | CS12 |
| impact fission | 0.05 | 0.90 | 0.01 | 2.40 | -5 | CS12 |
| impact fission | 0.05 | 0.93 | 0.02 | 2.31 | -5 | CS12 |
| Av. hit-and-run | 0.22±0.05 | 1.24±0.34 | 0.02±0.01 | 1.50±0.27 | -39±11 | |
| Av. merger | 0.43±0.03 | 1.15±0.21 | 0.02±0.01 | 2.05±0.21 | -3±24 ($|\delta f_T|$ = 19±13) | |
| Av. impact fission | 0.05±0.02 | 0.90±0.10 | 0.02±0.02 | 2.45±0.23 | -8±5 | |

*Successful simulation runs of the Giant Impact from the literature. For the impact fission model, the $\delta f_T$ value had to be calculated from related values given by Cúk & Stewart (2012). $M_{Theia}$ is the mass of the impacting body, $M_{Moon}$ is the mass of the final, accreted satellite, $Fe_{Disk}$ is the iron content of the disk, $L_{Final}$ is the final angular momentum of the Earth-Moon system. For all values, their averages and standard deviations (1σ) are given in the lowest three rows. Sources: R12 = Reufer et al., 2012; C12 = Canup, 2012; CS12 = Cúk & Stewart, 2012.*

**Table S3: Model parameters used for Figure 9**

| Parameter | Hit-and-run | | Merger | |
|---|---|---|---|---|
| | **Theia** | **Proto-Earth** | **Theia** | **Proto-Earth** |
| Mass ($M_E$) | 0.20 | 0.90 | 0.45 | 0.55 |
| $R_{surface}$ (km) | 4100 | 6200 | 5100 | 5400 |
| $G_{surface}$ (ms$^{-2}$) | 3.0 | 9.1 | 5.5 | 6.3 |
| $R_{core}$ (km) = $R_{CMB}$ | 2100 | 3300 | 2700 | 2900 |
| $R_{MO\,base}$ (km) | 700 | 1000 | 860 | 900 |
| $P_{MO\,base}$ (GPa) | 5.1 | 45 | 17 | 22 |
| $T_{MO\,base}$ (K) | 2100 | 3700 | 2600 | 2800 |
| av. $T_{MO\,base}$ (K)* | 2200 | 2800 | 2300 | 2400 |
| $\rho_{core}$ (gcm$^{-3}$) | 9 | 11 | 10 | 10 |
| $\delta^{30}$Si (bulk) | -0.62±0.05 | -0.42±0.05 | -0.62±0.05 | -0.42±0.05 |
| Si (%) | 23.5 | 20.6 | 23.5 | 19.0 |
| $f_{(T)}$ (%) | 53±3 | 87±6 | 55±1 | 55±1 |

*Average temperature at the magma ocean base throughout accretion.